\documentclass[aps,prl,twocolumn,showpacs,floatfix,nobibnotes,superscriptaddress,amsmath,amssymb,longbibliography]{revtex4-1}

\usepackage{graphicx}
\usepackage{epsfig}
\usepackage{bm}
\usepackage{color}
\usepackage{float}
\usepackage{dcolumn}
\usepackage{multirow} 
\usepackage{hyperref}
\usepackage{placeins}
\usepackage{multibib}
\usepackage{multirow}
\usepackage{lineno}

\newcites{app}{References}

\graphicspath{{.}{../}{figs/}}

\newcommand{\beq}{\begin{equation}}
\newcommand{\eeq}{\end{equation}}
\newcommand{\beqa}{\begin{eqnarray}}
\newcommand{\eeqa}{\end{eqnarray}}

\newcommand{\ket}[1]{\left| #1 \right\rangle}

\begin{document}
\begin{titlepage}

{\fontsize{26}{10} \textbf{\textcolor{black}{\flushleft Discrepancy between experimental and theoretical $\beta$-decay rates resolved from first principles}}}\\
{
 \\
P.~Gysbers$^{1,2}$, 
G.~Hagen$^{3,4}$, 
J.~D.~Holt$^{1}$,
G.~R.~Jansen$^{3,5}$,
T.~D.~Morris$^{3,4,6}$,
P.~Navr\'atil$^{1}$, 
T.~Papenbrock$^{3,4}$,
S.~Quaglioni$^{7}$,
A.~Schwenk$^{8,9,10}$,
S.~R.~Stroberg$^{1,11,12}$ \& 
K.~A.~Wendt$^{7}$
}

{
\fontsize{6}{10}{
\selectfont
$^{1}$TRIUMF, 4004 Wesbrook Mall, Vancouver BC, V6T 2A3, Canada.
$^{2}$Department of Physics and Astronomy, University of British Columbia, Vancouver BC, V6T 1Z1, Canada.
$^{3}$Physics Division, Oak Ridge National Laboratory, Oak Ridge, TN 37831, USA. 
$^{4}$Department of Physics and Astronomy, University of Tennessee, Knoxville, TN 37996, USA. 
$^{5}$National Center for Computational Sciences, Oak Ridge National Laboratory, Oak Ridge, TN 37831, USA.
$^{6}$Computational Sciences and Engineering Division, Oak Ridge National Laboratory, Oak Ridge, TN 37831, USA
$^{7}$Nuclear and Chemical Science Division, Lawrence Livermore National Laboratory, Livermore, CA 94551, USA.
$^{8}$Institut f\"{u}r Kernphysik, Technische Universit\"{a}t Darmstadt, 64289 Darmstadt, Germany.
$^{9}$ExtreMe Matter Institute EMMI, GSI Helmholtzzentrum f\"ur Schwerionenforschung GmbH, 64291 Darmstadt, Germany.
$^{10}$Max-Planck-Institut f\"ur Kernphysik, Saupfercheckweg 1, 69117 Heidelberg, Germany.
$^{11}$Physics Department, Reed College, Portland, OR 97202, USA. 
$^{12}$Department of Physics, University of Washington, Seattle, WA 98195, USA
}
}
\vspace{0.1cm}
\end{titlepage}
\textbf{$\beta$-decay, a process that changes a neutron into a proton
(and vice versa), is the dominant decay mode of atomic nuclei. This
decay offers a unique window to physics beyond the standard model, and
is at the heart of microphysical processes in stellar explosions and
the synthesis of the elements in the
Universe~\cite{janka2006,schatz2013,engel2017}. For 50 years, a
central puzzle has been that observed $\beta$-decay rates are
systematically smaller than theoretical predictions. This was
attributed to an apparent quenching of the fundamental coupling
constant $g_A \simeq$ 1.27 in the nucleus by a factor of about 0.75
compared to the $\beta$-decay of a free neutron~\cite{towner1987}. The
origin of this quenching is controversial and has so far eluded a
first-principles theoretical understanding. Here we address this
puzzle and show that this quenching arises to a large extent from the
coupling of the weak force to two nucleons as well as from strong
correlations in the nucleus. We present state-of-the-art computations
of $\beta$-decays from light and medium-mass nuclei to
$^{100}$Sn. Our results are consistent with experimental data,
including the pioneering measurement for
$^{100}$Sn~\cite{hinke2012,batist2010} (see Fig.~\ref{Sn100}). These
theoretical advances are enabled by systematic effective field
theories of the strong and weak interactions \cite{epelbaum2009}
combined with powerful quantum many-body techniques
\cite{barrett2013,hagen2015,stroberg2017}. This work paves the way for
systematic theoretical predictions for fundamental physics problems.
These include the synthesis of heavy elements in neutron star mergers
\cite{korobkin2012,surman2016,pian2017} and the search for
neutrino-less double-$\beta$-decay~\cite{engel2017}, where an
analogous quenching puzzle is a major source of uncertainty in
extracting the neutrino mass scale~\cite{barea2012}.}

\begin{figure}[h]
  \includegraphics[width=0.49\textwidth]{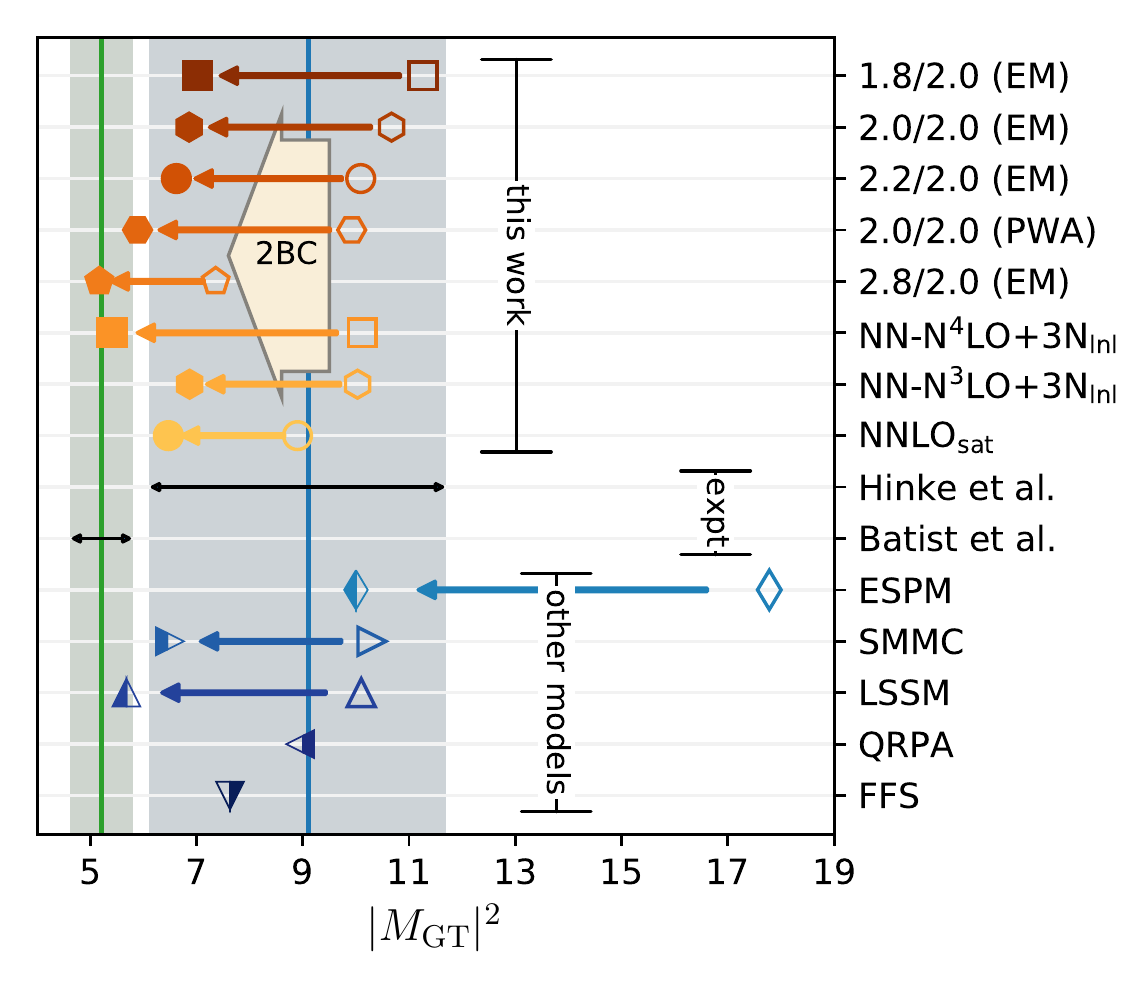}
  \caption{Gamow-Teller
    strength $|M_{\rm GT}|^2$ for the $\beta$-decay of $^{100}$Sn
    calculated in this work compared to data (``Hinke et
    al.''~\cite{hinke2012}), systematics (``Batist et
    al.''~\cite{batist2010}), and other models [extreme
      single-particle model (ESPM), shell-model Monte-Carlo (SMMC),
      large-space shell-model (LSSM), and finite Fermi systems (FFS)]
    from Ref.~\cite{hinke2012}. Hollow symbols represent results
    obtained with the standard Gamow-Teller operator
    ($\bm{\sigma\tau}$), full symbols also include two-body currents
    (2BC), and partially filled symbols show values following from the
    multiplication of the computed Gamow-Teller strength by the square
    of a phenomenological quenching factor. Each of our $^{100}$Sn calculations
    carry a conservatively estimated uncertainty of about 10\% (not shown).} \label{Sn100}
\end{figure}

Gamow-Teller transitions are a form of $\beta$-decay in which the
spins of the $\beta$-neutrino pair emitted during the nuclear decay
are aligned.  Remarkably, calculated Gamow-Teller strengths appear to
reproduce most of the experimental data if the fundamental constant
$g_A \simeq$ 1.27 characterizing the coupling of the weak interaction
to a nucleon is quenched by a factor of $q\sim
0.75$~\cite{wilkinson1973b,brown1985,chou1993,martinez1996}.  Missing
nuclear correlations (i.e. a lack of complexity in nuclear wave
functions owing to limitations of nuclear models) as well as neglected
contributions from meson-exchange currents (i.e. coupling of the weak
force to two nucleons) have been proposed as possible causes of the
quenching phenomenon~\cite{towner1987}. However, a solution has so far
remained elusive.  To address the quenching puzzle, we carry out a
comprehensive study of Gamow-Teller decays through many-body
computations of nuclei based on effective field theories (EFTs) of
quantum chromodynamics~\cite{epelbaum2009,machleidt2011}, including an
unprecedented amount of correlations in the nuclear wave
functions. The EFT approach offers the prospects of accuracy, by
encoding the excluded high-energy physics through coefficients
adjusted to data, and precision from the systematically improvable EFT
expansion.  Moreover, EFT enables a consistent description of the
coupling of weak interactions to two nucleons, via two-body currents
(2BC). In the EFT approach, 2BC enter as subleading corrections to the
one-body standard Gamow-Teller operator ${\bm \sigma} {\bm \tau}^+$
(with Pauli spin and isospin matrices ${\bm \sigma}$ and ${\bm \tau}$,
respectively); they are smaller but significant corrections to weak
transitions as three-nucleon forces are smaller but significant
corrections to the nuclear
interaction~\cite{epelbaum2009,machleidt2011}.

In this work we focus on strong Gamow-Teller transitions, where the
effects of quenching should dominate over cancellations due to fine
details (as occur in the famous case of the $^{14}$C decay used for
radiocarbon dating~\cite{holtjw2009,maris2011}).  An excellent example
is the super-allowed $\beta$-decay of the doubly magic $^{100}$Sn
nucleus (see Fig.~\ref{Sn100}), exhibiting the strongest Gamow-Teller
strength so far measured in all atomic nuclei~\cite{hinke2012}.  A
first-principles description of this exotic decay, in such a heavy
nucleus, presents a significant computational challenge. However, its
equal `magic' number ($Z=N=50$) of protons and neutrons arranged into
complete shells makes $^{100}$Sn an ideal candidate for large-scale
coupled-cluster calculations~\cite{morris2017}, while the daughter
nucleus $^{100}$In can be reached via novel extensions of high-order
charge-exchange coupled-cluster methods developed in this work (see
Methods and Figs.~\ref{fig:In100}, \ref{fig:Mgt_conv},
and~\ref{fig:Energy_conv} in Supplementary Information for
details). This method includes correlations via a vast number of
particle-hole excitations of a reference state and also employs 2BC in
the transition operator.

Figure~\ref{Sn100} shows our results for the strength (i.e., the
absolute square of the transition matrix element, $M_{\rm GT}$) of the
Gamow-Teller transition to the dominant $J^\pi=1^+$ state in the
$^{100}$In daughter nucleus (see Table~\ref{tab:tab2} and
Fig.~\ref{fig:Mgt_conv} in Supplementary Information for more
details). To investigate systematic trends and sensitivities to the
nuclear Hamiltonian, we employed a family of established EFT
interactions and corresponding currents~\cite{hebeler2011,
ekstrom2015, leistenschneider2017}. For an increased precision, we
also developed a new interaction labeled NN-N$^4$LO+3N$_{\rm lnl}$
which is constrained to reproduce the triton half-life (see Methods
for details on the Hamiltonians considered). The hollow symbols in
Fig.~\ref{Sn100} depict the decay with the standard, leading-order
coupling of the weak force to a single nucleon in the non-relativistic
limit (i.e., via the standard Gamow-Teller operator ${\bm \sigma}
{\bm \tau}^+$).  The differences with respect to the extreme
single-particle model (ESPM), which approximates both $^{100}$Sn and
its $^{100}$In daughter as a single shell-model configuration, reveals the
influence of correlations among the nucleons. The full symbols include
2BC, using consistent couplings as in the employed EFT
interactions. Finally, the partially filled symbols in
Fig.~\ref{Sn100} represent results from other models of
Ref.~\cite{hinke2012}, where the standard Gamow-Teller operator was
multiplied by a quenching factor $q \sim 0.75$.

Based on our results shown in Fig.~\ref{Sn100}, we predict the range
$5.2(5) \lesssim \vert M_{\rm GT}\vert^2 \lesssim 7.0(7)$ for the
Gamow-Teller strength. This range overlaps with the evaluation of
Batist {\it et al.}~\cite{batist2010}, based on systematic
experimental trends in the tin isotopes, and the lower end of the
measurement by Hinke {\it et al.}~\cite{hinke2012}. The quenching
factor we obtain from 2BC depends somewhat on the employed
Hamiltonian, and is in the range $q_{\rm 2BC} = 0.73-0.85$. This range
is consistent with the value $q=0.75(2)$ from Batist {\it et
al.}~\cite{batist2010}. In the present work, we used the spread of
results obtained with the selected set of EFT interactions and 2BC as
an estimate of the systematic uncertainty. A more thorough
quantification of the uncertainties associated with the many-body
methods and EFT truncations is beyond the scope of this work, and will
be addressed in future studies. We note that neglected higher-order
correlations in our coupled-cluster approach will further reduce the
Gamow-Teller strength (see Supplementary Information for details).

Moreover, we observe that the spread for the $^{100}$Sn Gamow-Teller
strength obtained for the family of employed EFT interactions is
significantly reduced (by a factor two) when 2BC are included. This is
consistent with ideas from EFT that residual cutoff dependence is due
to neglected higher-order terms in the Hamiltonian and 2BC. In
addition, we find that the relative contributions to the quenching of
the Gamow-Teller strength coming from correlations and 2BC vary as a
function of the resolution scale of the underlying EFT interactions.
Starting from the extreme single-particle model, and adding first
correlations and then the effects of 2BC, we find that the quenching
from correlations typically increases with increasing resolution scale
of the interaction, and that most of the quenching stems from
correlations. However, adding first the effects of 2BC and then
correlations shows that the quenching from 2BC increases with
decreasing resolution scale and that most of the quenching stems from
2BC for all but the ``hardest'' potentials considered in this work (see
Fig.~\ref{fig:sn100_espm} in Supplementary Information for
details).

\begin{figure}[t]
  \includegraphics[width=1.0\columnwidth]{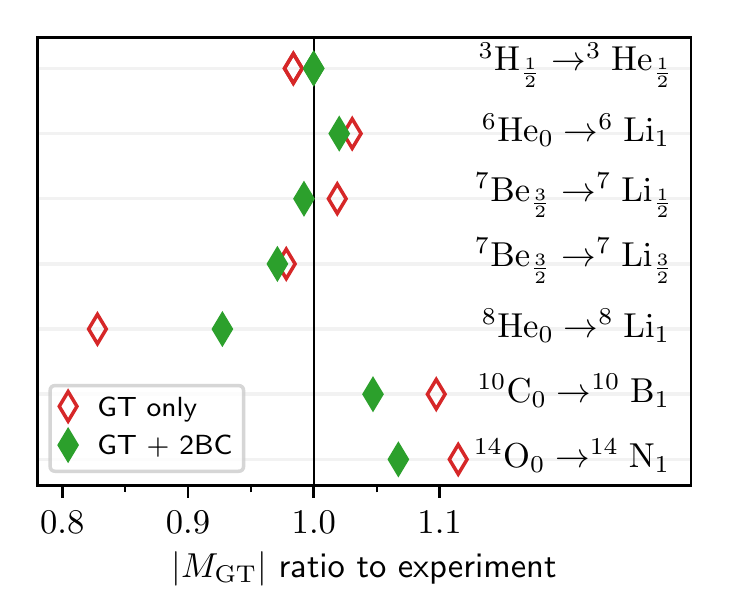}
  \caption{Theory-to-experiment ratio for the Gamow-Teller matrix
    elements of six strong transitions in light nuclei for the
    NN-N$^4$LO+3N$_{\rm lnl}$ interaction developed in this
    work. The subscripts in the legend denote the total angular
    momenta of the parent and daughter states. All initial states are
    ground states. In the case of $^3$H$\rightarrow^3$He,
    $^6$He$\rightarrow^6$Li and $^7$Be$\rightarrow
    ^7$Li$_{\tfrac{3}{2}}$, the daughter nucleus is in its ground
    state, while the $^7$Be$\rightarrow ^7$Li$_{\frac{1}{2}}$, $^{8}$He$\rightarrow ^{8}$Li$_{1}$  
    and $^{10}$C$\rightarrow ^{10}$B$_{1}$ are decays to the first excited
    state of the daughter nucleus, and the $^{14}$O$\rightarrow
    ^{14}$N$_{1}$ is a decay to the second excited state of
    $^{14}$N. Hollow symbols correspond to results obtained with the
    standard Gamow-Teller $\bm{\sigma\tau}$ operator, and full symbols include
    2BC.}
  \label{ncsm}
\end{figure}

For a comprehensive study, we now turn to $\beta$-decays of light and
medium-mass nuclei. Using a selection of the EFT interactions and 2BC
adopted for $^{100}$Sn, we achieved an overall good description of
$\beta$-decays in light nuclei.  Figure~\ref{ncsm} shows
theory-to-experiment ratios for large Gamow-Teller transitions in
light nuclei. Here we highlight the results obtained for the
high-precision NN-N$^4$LO+3N$_{\rm lnl}$ interaction and corresponding
2BC developed in this work. As detailed in Methods, the 2BC and
three-nucleon forces 3N$_{\rm lnl}$ are parametrized consistently and
are constrained to reproduce the empirical value of the triton
$\beta$-decay half life.  Our calculations were carried out with the
no-core shell model~\cite{barrett2013}, a virtually exact treatment of
correlations in the nuclear wave functions (see Methods for
details). The role of 2BC is relatively small in light nuclei with
mass numbers $A\leqslant 7$. Full nuclear wave functions already
provide a rather satisfactory description of the transitions with the
standard Gamow-Teller operator. Furthermore, the inclusion of 2BC may
enhance (e.g., $^8$He$\rightarrow^8$Li), quench (e.g.,
$^7$Be$_{\tfrac{3}{2}}\rightarrow^7$Li$_{\tfrac{1}{2}}$), or have
virtually no impact on the computed transition (e.g.,
$^7$Be$_{\tfrac{3}{2}}\rightarrow^7$Li$_{\tfrac{3}{2}}$, see also
Fig.~\ref{fig:Energy_conv4} in Supplementary Information). The small
role of 2BC in $A\leqslant 7$ nuclei is similar to what was found in
the Green's function Monte Carlo calculations of
Ref.~\cite{pastore2017}.  We find a rather substantial enhancement of
the $^8$He Gamow-Teller matrix element due to the 2BC. Let us mention,
though, that this transition matrix element is the smallest of those
presented in Fig.~\ref{ncsm}.  We note that for the other Hamiltonians
employed in this work, the 2BC and 3N were not fit to reproduce the
triton half-life, nevertheless the inclusion of 2BC for most of these
cases also improves the agreement with data for the light nuclei
considered in Fig.~\ref{ncsm} (see Fig.~\ref{ncsm2} in Supplementary
Information for results obtained with NNLO$_{\textrm{sat}}$ and
NN-N$^3$LO+3N$_{\rm lnl}$). The case of $^{10}$C is special because
the computed Gamow-Teller transition is very sensitive to the
structure of the $J^\pi=1^+$ state in the $^{10}$B daughter
nucleus. Depending on the employed interaction, this state can mix
with a higher-lying $1^+$ state, greatly impacting the precise value
of this transition. We finally note that benchmark calculations
between the many-body methods used in this work agree to within 5\%
for the large transition in $^{14}$O. For smaller transitions
discrepancies can be larger (see Supplementary Information for
details).

\begin{figure}[t]
  \includegraphics[width=1.0\columnwidth]{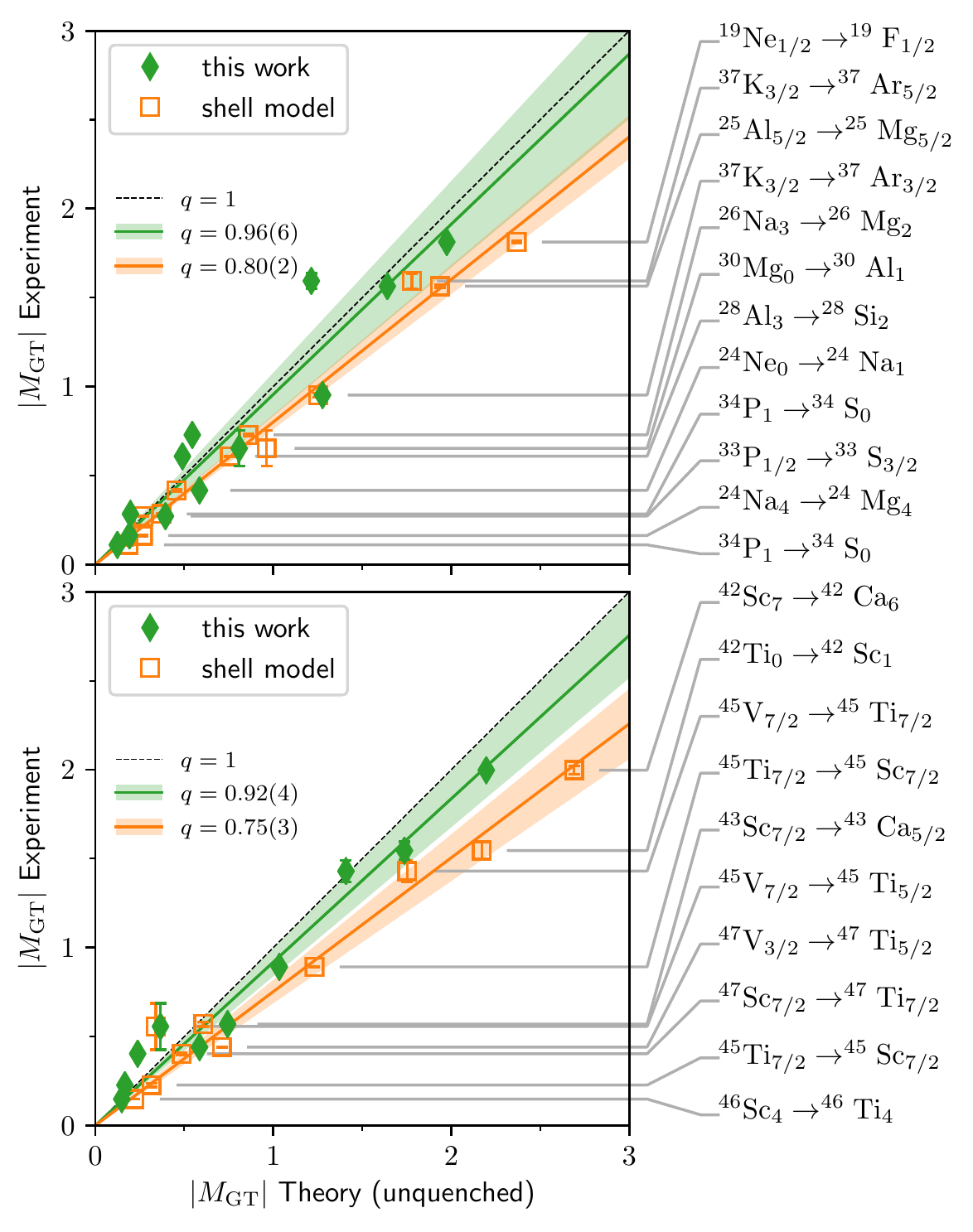}
  \caption{Comparison of experimental~\cite{ENSDF} and theoretical
    Gamow-Teller matrix elements for medium-mass nuclei in the
    $sd$-shell (top panel) and lower $pf$-shell (bottom panel). The
    theoretical results were obtained using phenomenological
    shell-model interactions~\cite{brown2006,martinez1996} with an
    unquenched standard Gamow-Teller ${\bm \sigma}{\bm \tau}$ operator 
    (orange squares); and
    using the VS-IMSRG approach with
    the NN-N$^4$LO+3N$_{\rm lnl}$ interaction and
    consistently evolved Gamow-Teller operator plus 2BC
    (green diamonds). The linear fits show the resulting quenching
    factor $q$ given in the panels, and shaded bands indicate
    one standard deviation from the average quenching factor.}
  \label{sdshell}
\end{figure}

Historically, the most extensive evidence for the quenching of Gamow-Teller
$\beta$-decay strength comes from medium-mass
nuclei~\cite{brown1985,langanke1995,martinez1996}, and we now show that our
calculations with these consistent Hamiltonians and currents largely
solves the puzzle here as well.  We use the valence-space in-medium
similarity renormalization group (VS-IMSRG) method~\cite{stroberg2017}
(see Methods for details) and compute Gamow-Teller decays for nuclei
in the mass range between oxygen and calcium (referred to as
$sd$-shell nuclei) and between calcium and vanadium (lower $pf$-shell
nuclei), focusing on strong transitions.  Here, we highlight the
NN-N$^4$LO+3N$_{\rm lnl}$ interaction and corresponding 2BC.

Figure~\ref{sdshell} shows the empirical values of the Gamow-Teller
transition matrix elements versus the corresponding unquenched
theoretical matrix elements obtained from phenomenological shell model
with the standard Gamow-Teller ${\bm \sigma}{\bm \tau}$ operator and
the first-principles VS-IMSRG calculations.  Perfect agreement between
theory and experiment is denoted by the diagonal dashed line.  The
results from the phenomenological shell model clearly exemplify the
state of theoretical calculations for
decades~\cite{wilkinson1973b,brown1985,chou1993,langanke1995,martinez1996}, as an
example, in the $sd$-shell shell, a quenching factor of $q\sim 0.8$ is
needed to bring the theory into agreement with
experiment~\cite{brown1985}. The VS-IMSRG calculations without 2BC
(not shown) exhibit a modest improvement, with a corresponding
quenching factor of 0.89(4) for $sd$-shell nuclei and 0.85(3) for
$pf$-shell nuclei, pointing to the importance of consistent
valence-space wave functions and operators (see~Fig.\ref{sdshell2} in
Supplementary Information). As in $^{100}$Sn, the inclusion of 2BC
yields an additional quenching of the theoretical matrix elements, and
the linear fit of our results lies close to the dashed line, meaning
our theoretical predictions agree on average with experimental values
across a large number of medium-mass nuclei.

Another approach often used in the investigation of Gamow-Teller
quenching is the Ikeda sum-rule: the difference between the total
integrated $\beta^-$ and $\beta^+$ strengths obtained with the
$ {\bm \sigma} {\bm \tau}^\mp $ operator yields the model-independent
sum-rule $3(N$ -- $Z)$.  We have computed the Ikeda sum-rule for
$^{14}$O, $^{48}$Ca, and $^{90}$Zr using the coupled-cluster method
(see Methods for details). For the family of EFT Hamiltonians used for
$^{100}$Sn we obtain a quenching factor arising from 2BC, which is
consistent with our results shown in Fig.~\ref{sdshell} and the
shell-model analyses from Refs.~\cite{chou1993,brown1985,langanke1995,martinez1996}
(see Fig.~\ref{Ikeda} in Supplementary Information).  We note that the
comparison with experimental sum-rule tests using charge-exchange
reactions~\cite{gaarde1981,wakasa1997} are complicated by the use of a
hadronic probe, which only corresponds to the leading weak one-body
operator, and by the challenge of extracting all strength to high
energies. Here, our developments enable future direct comparisons.

It is the combined proper treatment of strong nuclear correlations
with powerful quantum many-body solvers and the consistency between
2BC and three-nucleon forces that largely explains the quenching
puzzle. Smaller corrections are still expected to arise from neglected
higher order contributions to currents and Hamiltonians in the EFT
approach we pursued, and from neglected correlations in the nuclear
wave functions. For beyond-standard-model searches of new physics such
as neutrinoless double-$\beta$ decay, our work suggests that a
complete and consistent calculation without a phenomenological
quenching of the axial-vector coupling $g_A$ is called for. This
Letter opens the door to ab initio calculations of weak interactions
across the nuclear chart and in stars.

\bibliographystyle{naturemag_noURL}

\textbf{\textcolor{blue}{Acknowledgments}} We thank H. Grawe and
T. Faestermann for useful correspondance, J. Engel, E. Epelbaum,
D. Gazit, H. Krebs, D. Lubos, S. Pastore, and R. Schiavilla for useful
discussions, and K. Hebeler for providing us with matrix elements in
Jacobi coordinates for the three-nucleon interaction at
next-to-next-to-leading order~\cite{hebeler2011}.  This work was
prepared in part by Lawrence Livermore National Laboratory (LLNL)
under Contract DE-AC52-07NA27344 and was supported by the Office of
Nuclear Physics, U.S.\ Department of Energy, under Grants
DE-FG02-96ER40963, DE-FG02-97ER41014, DE-SC0008499, DE-SC0018223,
DE-SC0015376, the Field Work Proposals ERKBP57 and ERKBP72 at Oak
Ridge National Laboratory (ORNL), the FWP SCW1579, LDRD projects
18-ERD-008 and 18-ERD-058 and the Lawrence Fellowship Program at LLNL,
and by the NSERC Grant No.\ SAPIN-2016-00033, the ERC Grant No.~307986
STRONGINT, and the DFG under Grant SFB~1245.  TRIUMF receives federal
funding via a contribution agreement with the National Research
Council of Canada.  Computer time was provided by the Innovative and
Novel Computational Impact on Theory and Experiment (INCITE)
program. This research used resources of the Oak Ridge Leadership
Computing Facility located at ORNL, which is supported by the Office
of Science of the Department of Energy under Contract
No.\ DE-AC05-00OR22725. Computations were also performed at the LLNL
Livermore Computing under the institutional Computing Grand Challenge
Program, at Calcul Quebec, Westgrid and Compute Canada, and at the
J\"ulich Supercomputing Center (JURECA).\\

\textbf{\textcolor{blue}{Author contributions}} G.H., T.D.M., and
T.P.\ performed the coupled-cluster calculations. G.R.J. computed
three-nucleon forces for the coupled-cluster calculations. P.G., S.Q.,
P.N., and K.A.W.\ performed calculations for the two-body currents.
P.N.\ developed higher precision chiral three-nucleon interactions
used in this work and performed no-core shell model calculations.  G.H. and
T.D.M. derived and implemented new formalism to incorporate
higher-order excitations in coupled-cluster theory. S.R.S. and
J.D.H. performed VS-IMSRG calculations. All authors
discussed the results and contributed to the manuscript at all
stages.\\

\textbf{\textcolor{blue}{Author information}} 
Reprints and permissions information is available at www.nature.com/reprints.
The authors declare no competing financial interests. Readers are welcome to
comment on the online version of the paper. Correspondence and requests for
materials should be addressed to G.H. (hageng@ornl.gov)

\clearpage

\textbf{\textcolor{blue}{METHODS}}

\textbf{Hamiltonians and model space.} In this work we employ the
intrinsic Hamiltonian
\begin{equation}
  \label{intham}
  {H} = \sum_{i<j}\left(\frac{({\bf p}_i-{\bf p}_j)^2}{2mA} +
  {V}_{\rm NN}^{(i,j)}\right) + \sum_{ i<j<k}{V}_{\rm 3N}^{(i,j,k)}.
\end{equation}
\newcommand{\twoN}{{NN}}
\newcommand{\threeN}{{3N}}
Here ${\bf p}_i$ is the nucleon momentum, $m$ the average nucleon
mass, $A$ the mass number of the nucleus of interest, $V_{\rm NN}$ the
nucleon-nucleon (\twoN{}) interaction, and $V_{\rm 3N}$ the three-nucleon
(\threeN{}) interaction.

We use a set of interactions from Ref.~\cite{hebeler2011} labeled
1.8/2.0 (EM), 2.0/2.0 (EM), 2.2/2.0 (EM), 2.8/2.0 (EM), and 2.0/2.0
(PWA). These consist of a chiral \twoN{} interaction at order
N$^3$LO from Ref.~\citeapp{entem2003} evolved to the resolution scales
$\lambda_{\rm SRG} = 1.8,2.0,2.2,2.8$~fm$^{-1}$ by means of the
similarity renormalization group (SRG)~\citeapp{bogner2007} plus a
chiral \threeN{} interaction (unevolved) at order N$^2$LO, using a
non-local regulator with momentum cutoff
$\Lambda_{\rm 3N} = 2.0$~fm$^{-1}$. Note that the 2.0/2.0 (PWA)
interaction employs different long-range pion couplings in the \twoN{}
and \threeN{} sectors. The low-energy couplings entering these
interactions were adjusted to reproduce \twoN{} scattering data as
well as the $^3$H binding energy and $^4$He charge radius.
With the exception of 2.8/2.0 (EM), this set of interactions was
recently used to describe binding energies and spectra of neutron-rich
nuclei up to $^{78}$Ni~\citeapp{hagen2016b,simonis2017} and of
neutron-deficient nuclei around $^{100}$Sn~\cite{morris2017}.  The
results with the 1.8/2.0 (EM) interaction in particular reproduce
ground-state energies very well.

In addition, we also employ the NNLO$_{\rm sat}$ interaction, which
was constrained to reproduce nuclear binding energies and charge radii
of selected $p$- and $sd$-shell nuclei~\cite{ekstrom2015}.
\emph{Ab initio} calculations based on NNLO$_{\rm sat}$ accurately
describe both radii and binding energies of light- and medium-mass
nuclei~\cite{hagen2015},~\citeapp{lapoux2016,duguet2017}.

Finally, we employ two consistently SRG-evolved \twoN{} and \threeN{}
interactions, namely
NN-N$^3$LO+3N$_{\rm lnl}$~\cite{leistenschneider2017} and the
NN-N$^4$LO+3N$_{\rm lnl}$ introduced in this work. The \twoN{}
interactions at N$^3$LO and N$^4$LO are from Refs.~\citeapp{entem2003}
and \citeapp{entem2017}, respectively. The \threeN{} interactions
3N$_{\rm lnl}$ use a mixture of local~\citeapp{navratil2007} and
non-local regulators. The local cutoff is $650$ MeV, while the
nonlocal cutoff of $500$ MeV is the same as in the \twoN{}
interactions.  In case of the NN-N$^4$LO+3N$_{\rm lnl}$, the
parameters of the two-pion-exchange \threeN{} forces ($c_1$, $c_3$,
and $c_4$) are shifted with respect to their values in the \twoN{}
potential following the recommendation of
Ref.~\citeapp{entem2017}. The couplings of the shorter-range \threeN{}
forces ($c_D$ and $c_E$) are constrained to the binding energies and
radii of the triton and $^4$He in the NN-N$^3$LO+3N$_{\rm lnl}$ model,
and to the triton half-life and binding energy in the
NN-N$^4$LO+3N$_{\rm lnl}$ model. We note, however, that the
NN-N$^3$LO+3N$_{\rm lnl}$ interaction also reprodces the triton
half-life as shown in Fig.~\ref{ncsm2}.  The \twoN{} and \threeN{}
interactions are consistently SRG evolved to the lower cutoff
$\lambda_{\rm SRG} = 2.0\, \mathrm{fm}^{-1}$ (or
$\lambda_{\rm SRG} = 1.8\, \mathrm{fm}^{-1}$ in case of some of our
light nuclei calculations).

In our no-core shell model calculations of light nuclei we employ the
harmonic-oscillator basis varied in the range $N_{\rm max} =4-14$ and
with frequency $\hbar\omega = 20$~MeV. In our coupled-cluster and
valence-space IMSRG calculations we start from a Hartree-Fock basis
built from the harmonic-oscillator basis with model-space parameters
in the range $N_{\rm max}=6-14$ and $\hbar\omega = 12-16$~MeV,
respectively. Finally, the \threeN{} interaction is truncated
to three-particle energies with $E_{\rm 3max} \leqslant 16\hbar\omega$. \\

\textbf{Gamow-Teller transition operator.}
The rate at which a Gamow-Teller transition will occur is proportional
to the square of the reduced transition matrix element
\begin{equation}
  M_{\rm GT} = \langle f \vert {O}_{\rm GT}  \vert i \rangle\,.
\end{equation}
Here $i$ and $f$ label the initial and final states of the mother and
daughter nuclei, respectively.  (Note that throughout this work, we
quote the reduced matrix element $\langle f \| {O}_{\rm GT} \| i
\rangle$, using the Edmonds convention~\citeapp{edmonds1957}).  The
transition operator ${O}_{\rm GT}$ is defined in terms of the $J=1$
transverse electric multipole ${E}_{1}^{A}(\vec{K})$ of the
charge-changing axial-vector current $\vec{J}^A(\vec{K})$
\begin{equation}
  {O}_{\rm GT} = g_{\rm A}^{-1} \sqrt{6\pi}{E}_{1}^{A}(\vec{K} = 0)\,.
\end{equation}
Here, $\vec{K}$ denotes the momentum transferred to the resulting
electron and anti-neutrino pair (or positron and neutrino in
$\beta^+$-decay).  Because the change in energy between mother and
daughter states is typically very small (few MeV) compared to other
relevant scales, setting $|\vec{K}|=0$ is a very good approximation
that significantly simplifies the calculation of $M_{\rm GT}$.

\begin{figure}[h]
  \label{2bc}
  \includegraphics{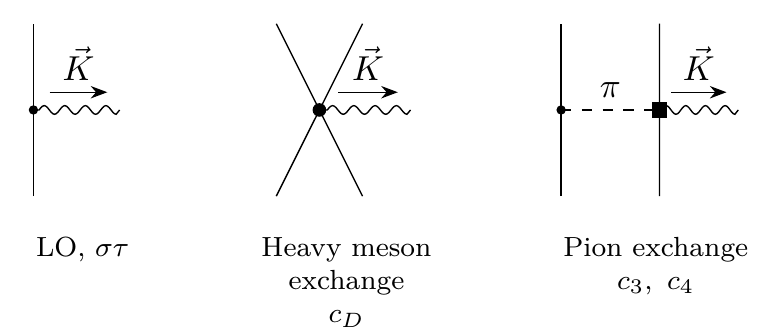}
  \caption{Diagrammatic
  representations of terms that contribute to the axial-vector
  charge-changing current up to N$^2$LO in the limit $\vec{K}\rightarrow0$, up
  to permutations of the vertices. The solid lines are nucleons, the dashed
  lines are pions, and the wavy line is the coupling to the electron and
  anti-neutrino. The leftmost diagram yields the standard (one-body) ${\bm \sigma}{\bm
  \tau}$ transition operator, the other diagrams form the leading 2BC.
  \label{fig:current} }
\end{figure}

The standard (one-body) charge-changing axial-vector current is
\begin{equation}
  \vec{J}^A(\vec{K}) = \sum_j \imath g_A \bm{\sigma}_j \bm{\tau}^{\pm}_j e^{\imath\vec{K}\cdot\vec{r}_j}\,.
  \label{eq:1b-current}
\end{equation}
Here $\vec{r}_j, \bm{\sigma}_j, \bm{\tau}^{\pm}_j $ are the position,
Pauli-spin, and charge-raising (lowering) operators for the $j$th
particle. In this work we use axial-vector currents derived within the
same chiral EFT framework used for the strong interactions, including
the leading 2BC~\citeapp{PhysRevC.67.055206}. The relevant
contributions (at $\vec{K}=0$) are depicted diagrammatically in
Fig.~\ref{fig:current}.  The leftmost diagram corresponds to the
leading-order one-body current. In addition, we include two classes of
2BC: a short-range term that shares a parameter ($c_D$) with the
one-pion-exchange \threeN{} force, as well as two long-range terms
that share parameters ($c_3$ and $c_4$) with the
two-pion-exchange \twoN{} and
\threeN{} forces. Within this framework, the Gamow-Teller operator
naturally decomposes into two major terms, the standard one-body
current (${O}_{\sigma\tau}^{1b} = \bm{\sigma\tau}^{\pm}$) and a
2BC,
\begin{equation} 
   {O}_{\rm GT} = {O}_{\sigma\tau}^{1b}+ {O}_{\rm 2BC}^{2b}.
  \label{gt-operator}
\end{equation}
 
For each of the chiral EFT Hamiltonians employed in this work, the
parameters $c_D$, $c_3$, and $c_4$ are taken consistently in the
\threeN{} force and 2BC, and the momentum cutoff for the
regularization of the currents, $\Lambda_{\rm 2BC}$, is set to the
value used in the non-local regulator of the \threeN{} interaction
(see Table \ref{tab:tab1} in Supplementary Information).  We found that the
choice of a local~\citeapp{gazit2009} versus non-local regulator in
the 2BC has a negligible effect on the Gamow-Teller transition
strength. The majority of our results were obtained using a local
regulator.  When appropriate, the currents were consistently evolved
with the nuclear forces to a lower resolution using the SRG (for
NN-N$^3$LO+3N$_{\rm lnl}$ and NN-N$^4$LO+3N$_{\rm lnl}$), keeping only
up to two-body contributions. In light nuclei, three- and higher-body
SRG-induced
terms are very small. \\

\textbf{Quantum many-body methods.} In what follows we describe the
many-body methods used in this work: coupled-cluster theory, the
no-core shell model, and the IMSRG. The coupled-cluster
calculations required new methodological developments, which are
described in detail
below. \\

\textbf{Coupled-cluster method.} Our coupled-cluster
calculations start from a Hamiltonian $H_N$ that is normal-ordered 
with respect to a single-reference Hartree-Fock state $\vert
\Phi_0\rangle$. We approximate the full 3N interaction by
truncating it at the normal-ordered two-body level. This approximation
has been shown to work well for light- and medium-mass nuclei
\citeapp{hagen2007a,roth2012,hergert2013b}. The central quantity in the
coupled-cluster method is the similarity transformed Hamiltonian
$\overline{H}_N = e^{-T} H_N e^T$, with $T=T_1 + T_2 + \ldots$ being a
linear expansion of particle-hole excitations with respect to the
reference state $\vert \Phi_0\rangle$. The truncation of this expansion at
some low-order particle-hole excitation rank is the only approximation
that occurs in the coupled-cluster
method~\citeapp{bartlett2007,hagen2014}. The non-Hermitian
Hamiltonian $\overline{H}_N$ is correlated and the reference state
$\vert \Phi_0\rangle$ becomes the exact ground-state.

We compute ground and excited states using the CCSDT-1 and EOM-CCSDT-1
approximations~\citeapp{lee1984,watts1995}, respectively. These
approximations include iterative singles and doubles and
leading-order triples excitations, and capture about 99\% of the
correlation energy in closed (sub-) shell
systems~\citeapp{bartlett2007}.
For the Gamow-Teller transitions and expectation
values, we solve for the left ground-state of $\overline{H}_N$ 
\begin{equation}
\label{check}
\langle \Phi_0 \vert (1+\Lambda), \quad {\rm with} \quad \Lambda= \Lambda_1 + \Lambda_2
+ \ldots .
\end{equation}
Here $\Lambda $ is a linear expansion in particle-hole de-excitation
operators. We truncate $\Lambda $ at the EOM-CCSDT-1
level consistent with the right CCSDT-1 ground-state solution~\citeapp{watts1995}. 

The Gamow-Teller transition of a $J^\pi=0^+$ ground state occupies
low-lying $1^+$ states in the daughter nucleus. These states in the
daughter nucleus are calculated by employing the charge-exchange
equation-of-motion coupled-cluster method~\citeapp{ekstrom2014}, and
we also include the leading-order three-particle-three-hole
excitations as defined by the EOM-CCSDT-1
approximation~\citeapp{watts1995}. The absolute squared Gamow-Teller
transition matrix element is then
\begin{eqnarray}
  \nonumber
  \vert M_{\rm GT}\vert^2 = \vert \langle f \vert {O}_{\rm GT} \vert
  i \rangle\vert^2 = \langle f \vert {O}_{\rm GT} \vert i \rangle
  \langle i \vert {O}_{\rm GT}^\dagger \vert f \rangle  \\
  = \langle \Phi_0 \vert L_\mu^{1^+} \overline{ O_{N}} \vert
  \Phi_0\rangle \langle \Phi_0 \vert (1+\Lambda) \overline{O_{N}^\dagger} R_{\mu}^{1^+}\vert \Phi_0 \rangle. 
\end{eqnarray}
Here $R_\mu^{1^+} $ is the right and $L_\mu^{1^+} $ the corresponding
left excited $1^+$ state in the daughter nucleus, and
$\overline{O_N} = e^{-T} {O}_{N} e^T$ is the similarity transform of the
normal-ordered Gamow-Teller operator ${O}_{\rm GT}$ [see
Eq.~\eqref{gt-operator}]. In ${O}_N$ we approximate the two-body part of
the operator ${O}_{\rm GT}$ at the normal-ordered one-body level,
neglecting the residual two-body normal-ordered
part~\citeapp{ekstrom2014}.  Note that the construction of
$\overline{O}_N$ induces higher-body terms, and we truncate
$\overline{O}_N$ at the two-body level. This approximation is precise
for the case of electromagnetic sum rules in
coupled-cluster theory \citeapp{miorelli2018}.

We evaluate the total integrated Gamow-Teller strengths as a
ground-state expectation value
\begin{equation} 
S^{\pm} = \langle \Phi_0\vert (1+\Lambda) \overline{O^\dagger_N}
\cdot \overline{O_N}\vert \Phi_0 \rangle .
\end{equation}
For ${O} = {O}_{\sigma\tau}^{1b}$ the Ikeda sum-rule is $S^- -
S^+ = 3(N - Z)$. As a check of our code, we have verified that this
sum-rule is fulfilled.

The inclusion of triples excitations of the right and left eigenstates
$R_\mu$ and $L_\mu $, respectively, is challenging in terms of CPU
time and the memory.  To limit CPU time, we restrict the employed
three-particle--three-hole configurations in the EOM-CCSDT-1
calculations to the vicinity of the Fermi surface. This is done by
introducing a single-particle index $\tilde{e}_p=|N_p-N_{F}|$ that
measures the difference between the numbers of oscillator shells $N_p$
of the single-particle state with respect to the Fermi surface
$N_F$. We only allow three-particle and three-hole configurations with
$\tilde{E}_{pqr}=\tilde{e}_p+\tilde{e}_q+\tilde{e}_r<\tilde{E}_{\rm
  3max}$. This approach yields a rapid convergence in EOM-CCSDT-1
calculations, as seen in Fig. \ref{fig:Mgt_conv} of the Supplementary
Information.

The storage of the included three-particle--three-hole amplitudes
exceeds currently available resources and had to be avoided. We follow
Refs.~\citeapp{bloch1958,haxton2000,smith2005} and define an effective
Hamiltonian in the $P$ space of singles and doubles excitations, so
that no explicit triples amplitudes need be stored. Denoting the
$Q$-space as that of all triples excitations below $\tilde{E}_{\rm
3max}$, the right eigenvalue
equation can be rewritten as
\begin{equation}
\begin{bmatrix} 
\overline{H}_{PP} & \overline{H}_{PQ} \\
\overline{H}_{QP} & \overline{H}_{QQ} 
\end{bmatrix}
\begin{bmatrix} 
R_{P}  \\
R_{Q} 
\end{bmatrix}
=\omega
\begin{bmatrix} 
R_{P}  \\
R_{Q} 
\end{bmatrix}\\ .
\end{equation} 
This yields 
\begin{equation}
\overline{H}_{PP}R_P + \overline{H}_{PQ}R_Q = \omega R_P , 
\label{eq:pspace}
\end{equation}
and 
\begin{equation}
\overline{H}_{QP}R_P + \overline{H}_{QQ}R_Q = \omega R_Q \\ .
\label{eq:qspace}
\end{equation}
Here we have suppressed the label $\mu$ denoting different excited
states.  Solving Eq.~(\ref{eq:qspace}) for the triples component of $R$,
and then substituting into Eq.~(\ref{eq:pspace}), we arrive at
\begin{equation}
\overline{H}_{PP}R_P + \overline{H}_{PQ}(\omega-\overline{H}_{QQ})^{-1}\overline{H}_{QP}R_P = \omega R_P.
\label{eq:BH_eq}
\end{equation}
In the EOM-CCSDT-1 approximation
$\overline{H}_{QQ}=\langle T|{F}|T\rangle$, where ${F}$ is the
Fock matrix. In the Hartree-Fock basis $\overline{H}_{QQ}$ is
diagonal, and its inversion is trivial.  We solve this
energy-dependent, effective Hamiltonian self-consistently to arrive at
exact eigenstates of the EOM-CCSDT-1 Hamiltonian. This allows for only
one state to be constructed at a time.  For the computation of higher
spin excited states in the daughter nucleus $^{100}$In, we combine the
iterative EOM-CCSDT-1 approach, with a perturbative approach that
accounts for all excluded three-particle-three-excitations outside the
energy cut $\tilde{E}_{\rm 3max}$. This approach is analogous to the
active space coupled cluster methods of
\citeapp{shen2012a,shen2012b}. By denoting the $Q'$-space as that of
all three-particle-three-hole excitations above
$\tilde{E}_{\rm 3max}$, we arrive at the following perturbative
non-iterative energy correction,
\begin{equation}
\Delta \omega_{\mu} = \langle \Phi_0 \vert L_\mu \overline{H}_{PQ'}(\omega_\mu-\overline{H}_{Q'Q'})^{-1}\overline{H}_{Q'P}R_\mu \vert \Phi_0 \rangle  \, .
\label{eq:qecorr}
\end{equation}
Here $R_\mu$ and $L_\mu $ are the right and corresponding left
EOM-CCSDT-1 eigenstates obtained from diagonalization of the
energy-dependent similarity transformed Hamiltonian given in
Eq.~(\ref{eq:BH_eq}).  We label this approach EOM-CCSDt-1, and it
drastically improves convergence to the full-space EOM-CCSDT-1
energies (see Fig.~\ref{fig:Energy_conv} in Supplementary Information
for details).
\\

\textbf{No-core shell-model.}  The no-core shell model
(NCSM)~\cite{barrett2013}, \citeapp{navratil2000} treats nuclei as
systems of $A$ non-relativistic point-like nucleons interacting
through realistic inter-nucleon interactions. All nucleons are active
degrees of freedom. The many-body wave function is cast into an
expansion over a complete set of antisymmetric $A$-nucleon
harmonic-oscillator basis states containing up to $N_{\rm max}$
harmonic-oscillator excitations above the lowest
Pauli-principle-allowed configuration:
\begin{equation}\label{NCSM_wav}
 \ket{\Psi^{J^\pi T}_A} = \sum_{N=0}^{N_{\rm max}}\sum_i c_{Ni}^{J^\pi T}\ket{ANiJ^\pi T}\; .
\end{equation}
Here, $N$ denotes the total number of harmonic-oscillator excitations
of all nucleons above the minimum configuration, $J^\pi T$ are the
total angular momentum, parity, and isospin, and $i$ denotes
additional quantum numbers. The sum over $N$ is restricted by parity
to either an even or odd sequence. The basis is further characterized
by the frequency $\omega$ of the harmonic
oscillator. Square-integrable energy eigenstates are obtained by
diagonalizing the intrinsic Hamiltonian typically by applying the
Lanczos algorithm. In the present work, we used the
importance-truncation NCSM~\citeapp{roth2008a} to reduce the basis
size in the highest $N_{\rm max}$ spaces of the $A=10$ and $A=14$
nucleus calculations. \\

\textbf{Valence-space in-medium similarity renormalization group.}
The in-medium similarity renormalization group
(IMSRG) \citeapp{tsukiyama2011,tsukiyama2012,hergert2016} transforms
the many-body Hamiltonian ${H}$ to a diagonal or block-diagonal form via
a unitary transformation ${U}$, i.e., it generates
$\tilde{H}={U}{H}{U}^{\dagger}$. To achieve this, one expresses the
transformation as the exponential of an anti-Hermitian generator,
${U}=e^{{\Omega}}$. Here ${\Omega}$ encodes information on the off-diagonal
physics to be decoupled~\citeapp{morris2015}. Beginning
from some single-reference ground-state configuration
$| \Phi_0 \rangle$ (e.g., the Hartree-Fock state based on initial
interactions), we map the reference to the fully correlated ground
state $|\Psi_0 \rangle$ via a continuous sequence of such unitary
transformations ${U}(s)$. With no approximations, this gives the exact
ground-state energy, but in the IMSRG(2)
approximation used here, all operators are truncated at the two-body
level.

In the valence-space formulation,
VS-IMSRG~\citeapp{bogner2014,stroberg2017}, the unitary transformation
is constructed (based on a redefinition of ${\Omega}$) to in addition
decouple a valence-space Hamiltonian ${H}_{\rm vs}$ from the remainder
of the Hilbert space.  We use an ensemble
reference~\citeapp{stroberg2017} state for normal ordering to capture
the main effects of three-body operators within the valence space.
The eigenstates are obtained by a subsequent diagonalization of
${H}_{\rm vs}$ within the valence space.  Furthermore, any general
operator ${\mathcal{O}}$ can then be transformed by
$\tilde{\mathcal{O}}=e^{{\Omega}} {\mathcal{O}} e^{- {\Omega}}$, to
produce an effective valence-space operator consistent with the
valence-space Hamiltonian~\citeapp{parzuchowski2017}. The expectation
value of ${\mathcal{O}}$ between initial and final states is obtained
as usual by combining the matrix elements of ${\mathcal{O}}$ with the
one- and two-body shell-model transition densities.  Note that there
is some ambiguity about which reference we should take when normal
ordering: the parent or the daughter.  If we were able to perform the
unitary transformation without approximation, either choice should
give exactly the same answer, so long as we use the same
transformation on the wave functions and the operators.  However,
because we truncate at the two-body level, the transformation is not
unitary and the error made is reference-dependent.  Comparing results
obtained by normal ordering with respect to the parent or the daughter
nucleus then provides a (lower bound) estimate of the error due to the
truncation.  In this work, we find that the different choices give
transition matrix elements that differ on the order of $\sim$~5\%.
The results presented are those obtained with the parent as the
reference.  (As an example, if we use $^{14}$N as the reference, the
numbers in the third line of Table~\ref{tab:tab4} become 1.77, 1.81,
1.88, and 1.87).

\bibliographystyleapp{naturemag_noURL}

\clearpage

\title{\textcolor{blue}{Supplementary Information}}
\maketitle

\subsection{Spectrum of $^{100}$In}

\begin{figure}[h] 
  \includegraphics[width=1.0\columnwidth]{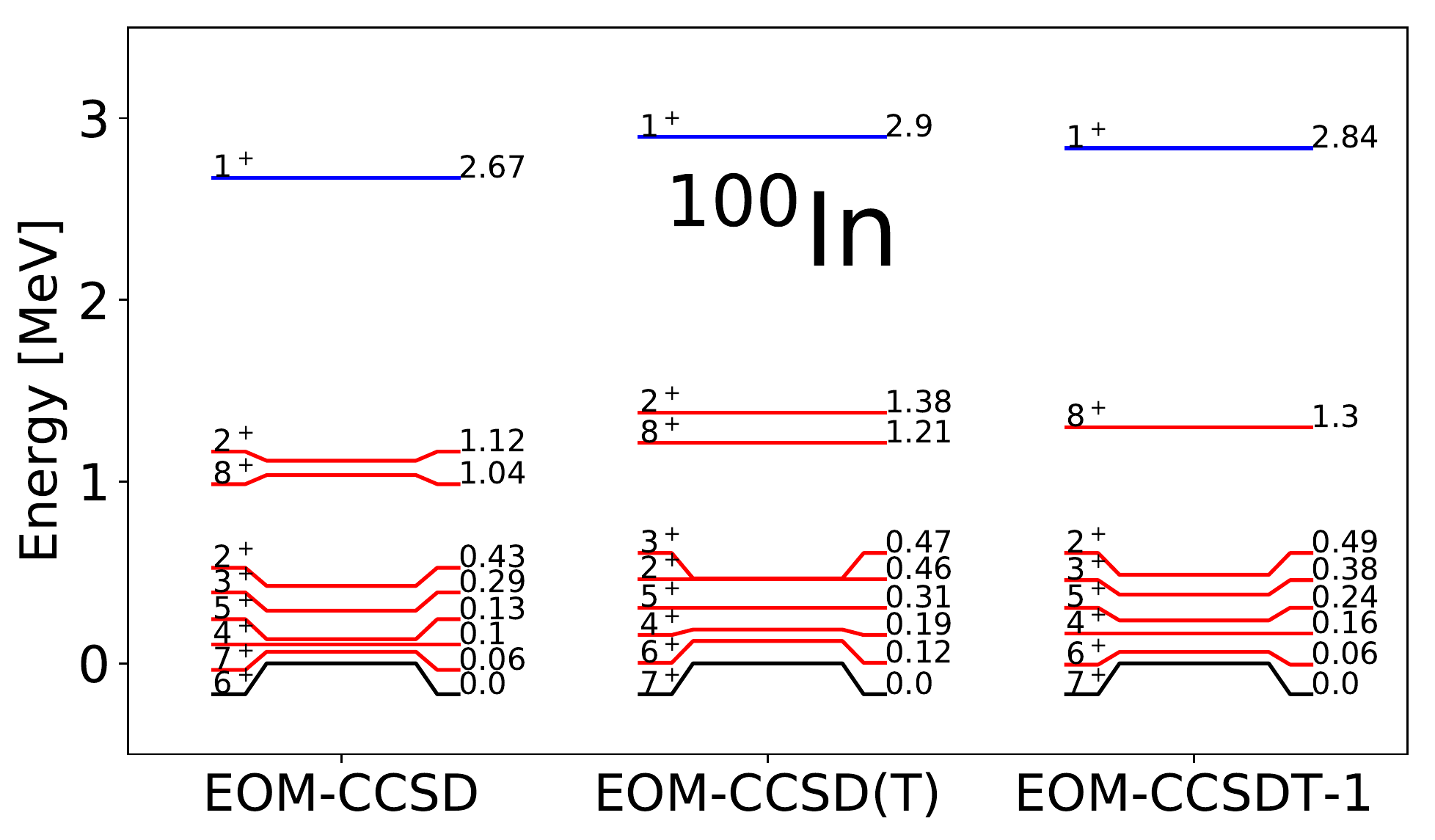} \caption{(Color
    online) Spectrum of $^{100}$In obtained with the 1.8/2.0 (EM)
    interaction computed using the EOM-CCSD, EOM-CCSD(T) and
    EOM-CCSDT-1 methods. The spins/parities and excitation energies
    (in MeV) of low-lying states and of the $1^+$ state (blue level) populated in the
    $\beta$ decay of $^{100}$Sn are shown.} \label{fig:In100}
\end{figure}

Let us present details regarding the quality of our caluclation of
$^{100}$In, the daughter nucleus of the Gamow-Teller decay of
$^{100}$Sn. Unfortunately, only little is known regarding the
structure of this nucleus~\cite{hinke2012}.  Figure~\ref{fig:In100}
shows the spectrum of the daughter nucleus $^{100}$In for the 1.8/2.0
(EM) interaction, computed with the EOM-CCSD, EOM-CCSD(T), and
EOM-CCSDT-1 methods. While triples (T) correlations only change the
excitation energies of quasi-degenerate states, they add about 1~MeV
of binding energy to $^{100}$In. Due to the large level density in
this odd-odd nucleus, it is difficult to predict the (unknown)
ground-state spin.  The $\beta$-decay of $^{100}$Sn populates $1^+$
states in the daughter. The excitation energy $E_{1^+}\approx 2.9$~MeV
is consistent with data~\cite{hinke2012}. Overall, the spectrum agrees
also with large-scale shell model
calculations~\citeapp{faestermann2013}.

\subsection{Role of correlations and 2BC in the Gamow-Teller decay of $^{100}$Sn}

\begin{figure}[h]
  \includegraphics[width=1.0\columnwidth]{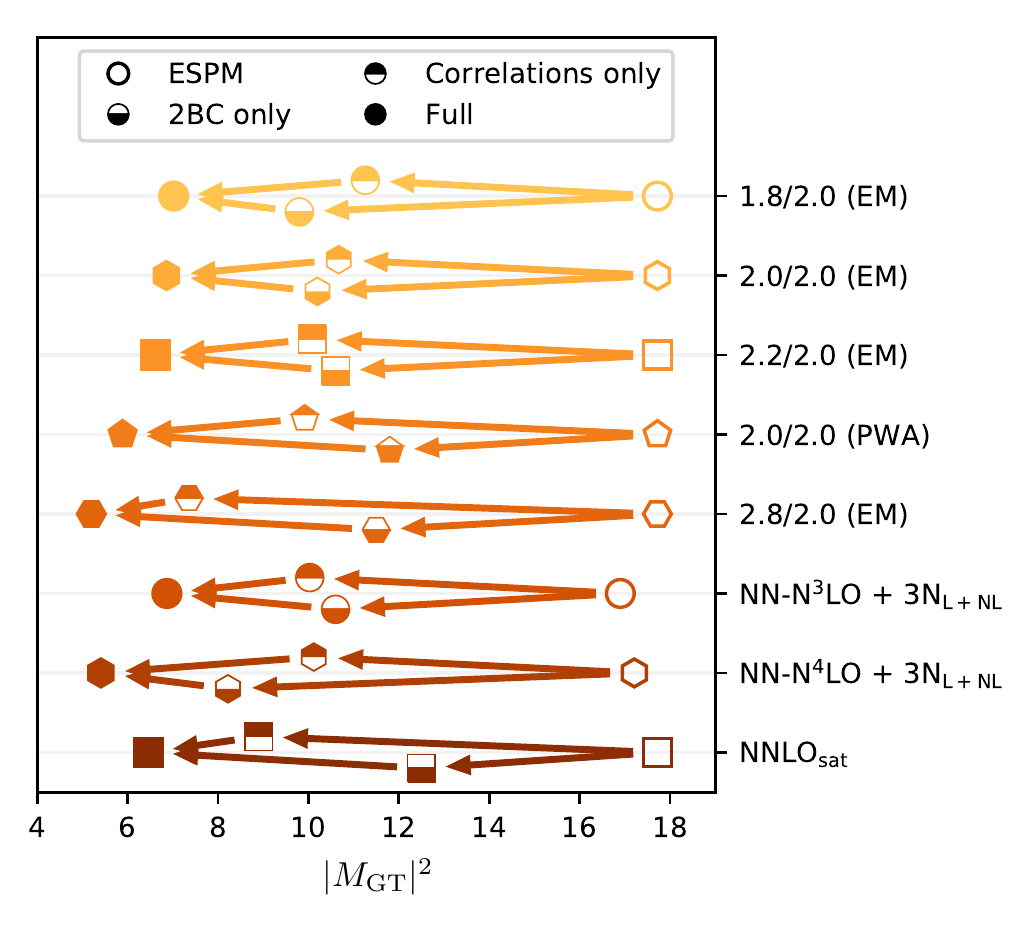} \caption{(Color
  online) Comparison of the role of correlations versus 2BC for the
  family of EFT interactions employed in this work. Here the hollow
  symbols show the single-particle transition matrix element (ESPM) from the
  $\nu g_{9/2}$ to the $\pi g_{7/2}$ orbitals. The partially filled
  symbols show the the fully correlated transition matrix element
  with the standard Gamow-Teller operator only (upper-half filled), and the
  single-particle transition matrix element from the $\nu g_{9/2}$ to
  the $\pi g_{7/2}$ orbitals including 2BC (lower-half filled). The filled symbols show
  the final correlated Gamow-Teller transition matrix element with 2BC
  included.} \label{fig:sn100_espm}
\end{figure}

To throw light onto the quenching puzzle of Gamow-Teller decays, we
analyse the contributions of 2BC and strong many-body correlations.
Figure~\ref{fig:sn100_espm} analyzes the role of strong correlations
and 2BC for the family of EFT interactions employed in this work. This
analysis can be done in two ways. First, we can compare the
single-particle transition matrix element (ESPM) from the $\nu
g_{9/2}$ to the $\pi g_{7/2}$ orbital using the standard Gamow-Teller
operator to that with 2BC included, and then finally with the matrix
element where correlations are included. This is shown as the lower
paths in Fig.~\ref{fig:sn100_espm}. Second, we can compare the
single-particle transition matrix element from the $\nu g_{9/2}$ to
the $\pi g_{7/2}$ orbital using the standard Gamow-Teller operator to
the fully correlated transition matrix element using the standard
Gamow-Teller operator only ($\vert M_{\rm
  GT}(\bm{\sigma\tau})\vert^2$), and then to the fully correlated
transition matrix element with 2BC included ($\vert M_{\rm
  GT}\vert^2$) as shown in the upper paths. Depending whether one goes
along the upper or lower path the role of correlations versus the role
of 2BC on the quenching is different. Of course, only the sum of the
effects from correlations and 2BC are observable. This analysis shows
that the interplay of correlations and 2BC is subtler than often
portrayed (when almost always the upper path is
considered)~\citeapp{brown1988,gazit2009,ekstrom2014},\cite{pastore2017}.
We also note that the contribution from correlations (2BC) to the
quenching typically increases (decreases) with increasing resolution
scale (i.e. increasing ultraviolet cutoff) of the employed
interaction. This is not unexpected, as ``hard'' interactions
introduce more correlations into wave functions and thereby reduce the
large transition matrix element one would expect from a transition
between two shell-model configurations.

\subsection{Summary of results for $^{100}$Sn}

\begin{table*}[hbt]
  \caption{Gamow Teller (GT) matrix elements from the extreme
    single-particle model (ESPM) $\vert m_{\rm GT} (\bm{\sigma\tau})
    \vert^2 $ and EOM-CCSDT-1 $\vert M_{\rm GT} (\bm{\sigma\tau}) \vert^2
    $, sum of GT and 2BC from ESPM $\vert m_{\rm GT}\vert^2 $ and
    EOM-CCSDT-1 $\vert M_{\rm GT}\vert^2 $, the log $ft$ values, the quenching factor $q$
    from 2BC from EOM-CCSDT-1 and ESPM, the energy difference between
    the ground-state of $^{100}$Sn and the $1^+$ state in $^{100}$In,
    and the binding energy per particle ($BE/A$) in $^{100}$Sn for the
    interactions used in this work, and compared to experiment. The
    $BE/A$ obtained with the interactions X/Y (EM) and 2.0/2.0
    (PWA) were taken from Ref.~\protect\cite{morris2017}, the $BE/A$
    for the remaining interactions were obtained by employing the same
    model-space as used for the calculations of $\vert M_{\rm
      GT}\vert^2 $, i.e., $N_{\rm max} = 10$ and $E_{\rm 3max} =
    16$. The energy difference $\Delta E$ were calculated using the
    model-space $N_{\rm max} = 10$ and $E_{\rm 3max} = 16$ for the
    ground-state (CCSDT-1) and $\tilde{E}_{\rm 3max} = 11$ for the
    $1^+$ excited state (EOM-CCSDT-1) in $^{100}$In. All calculations
    used the harmonic oscillator frequency $\hbar\omega = 12$MeV.}
\begin{center}
\renewcommand{\arraystretch}{1.3}
\begin{tabular}{|l|c|d|c|l|l|c|c|l|l|}\hline
 Interaction & $\vert m_{\rm GT} (\bm{\sigma\tau}) \vert^2 $ & \vert M_{\rm GT} (\bm{\sigma\tau}) \vert^2 &
 $\vert m_{\rm GT}\vert^2 $  & $\vert M_{\rm GT}\vert^2$  & log $ft$ & {\it q} &  {\it q} (ESPM) & $\Delta E$ [MeV] & $BE/A$ [MeV] \\\hline
 NNLO$_{\rm sat}$           & 17.7 & 8.9     & 12.5 &  6.5  &  2.77  & 0.85 & 0.84 & 7.4 &  not converged \\
 NN-N$^3$LO+3N$_{\rm lnl}$& 16.9 & 10.0 & 10.6& 6.9  & 2.74  & 0.83 & 0.79 & 6.1 & 7.6 \\
 NN-N$^4$LO+3N$_{\rm lnl}$& 17.2 & 10.1 & 8.2& 5.4  & 2.85 &  0.73 & 0.69 & 5.8 & 7.1 \\
 1.8/2.0 (EM)               & 17.7    & 11.3 & 9.8 & 7.0 &  2.73 & 0.79 & 0.74 & 5.1 & 8.4 \\
 2.0/2.0 (EM)               & 17.7    & 10.7 & 10.2 & 6.9 &  2.74 & 0.80 & 0.76 & 6.0 & 7.7  \\
 2.0/2.0 (PWA)             & 17.7    & 9.9   & 11.5 & 5.9 &  2.81 & 0.77 & 0.81 & 6.8 & 6.4  \\
2.2/2.0 (EM)                & 17.7    & 10.1 & 10.6 & 6.6 &  2.76 & 0.81 & 0.77 & 6.7 &  7.2 \\
2.8/2.0 (EM)                & 17.7    & 7.4   & 11.8 & 5.2 & 2.86 & 0.84 & 0.82& 8.3 &  not converged \\\hline
 Batist {\it et al.}~\cite{batist2010} &\multirow{2}{*}{}
 &\multirow{2}{*}{}&\multirow{2}{*}{}& $5.2 \pm 0.6$ &\multirow{2}{*}{}
 &\multirow{2}{*}{}  & & \multirow{2}{*}{5.11} & \multirow{2}{*}{8.25} \\
 Hinke {\it et al.}~\cite{hinke2012} & & & & $9.1^{+2.6}_{-3.0}$& $2.62^{+0.13}_{-0.11}$ &  & & &  \\\hline
\end{tabular}
\renewcommand{\arraystretch}{1}
\end{center}
\label{tab:tab2}
\end{table*}

Let us summarize our extensive results for the key heavy nucleus
$^{100}$Sn.  Table~\ref{tab:tab2} shows the results for the $\beta^+$
decay of $^{100}$Sn to $^{100}$In for the various Hamiltonians
employed in this work. The listed observables are the Gamow-Teller
strenghts with 2BC ($\vert M_{\rm GT}\vert^2$), the log~$ft$ values,
the Gamow-Teller strengths with the standard Gamow-Teller operator
only ($\vert M_{\rm GT}(\bm{\sigma\tau})\vert^2 $), the energy
difference between the ground state of $^{100}$Sn and the first
excited $1^+$ state in the daughter nucleus $^{100}$In, and the
binding energy per nucleon ($BE/A$) for the family of EFT interactions
employed in this work. For the cases where the binding energy is not
converged, we note that the Gamow-Teller operator acts only in
spin-isospin space, and is therefore rather insensitive to the tail of
the wave functions (which carry information about the binding
energy). Benchmark calculations between EOM-CCSDT-1 and NCSM for the
Gamow-Teller transition in $^{14}$O show agreement with 3\% for three
different interactions (see Table~\ref{tab:tab4} below). Therefore a
conservative uncertainty estimate for the Gamow-Teller transition in
$^{100}$Sn is about 10\%.

\subsection{Results for Ikeda sum-rule}

\begin{figure}[ht]
  \includegraphics[width=1.0\columnwidth]{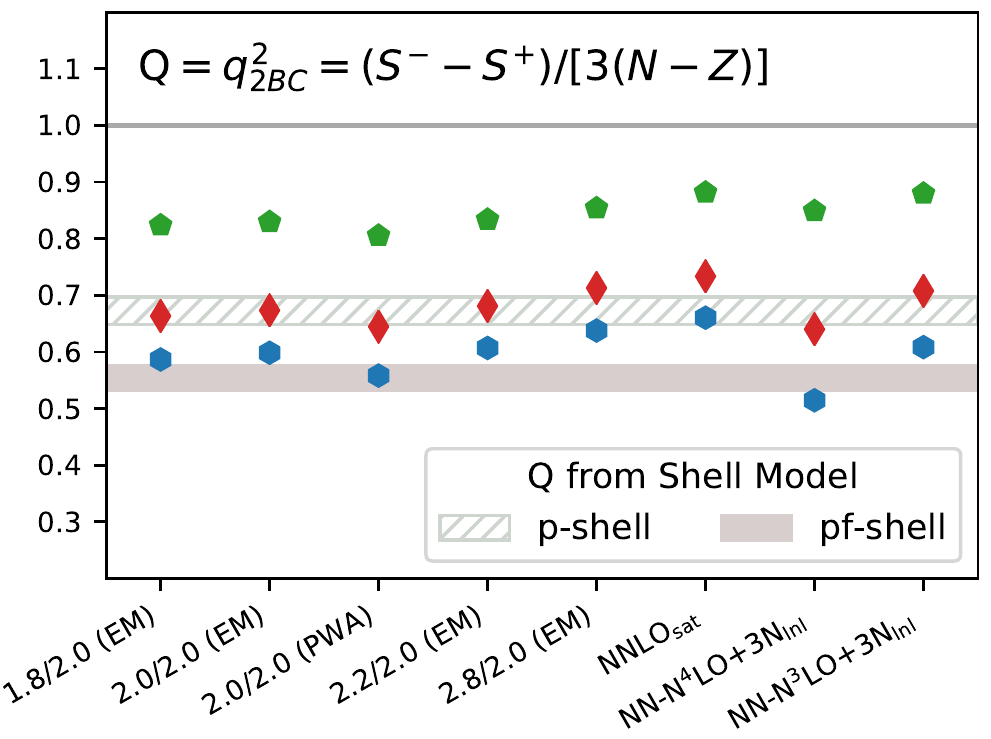} \caption{
    Predicted quenching, $Q \equiv q_{\rm 2BC}^2 = (S^- - S^+)/ [3(N -
      Z)]$, of the Ikeda sum-rule in $^{14}$O (green filled
    pentagons), $^{48}$Ca (red filled diamonds), and $^{90}$Zr (blue
    filled hexagons) due to 2BC. The quenching factor is calculated
    for the family of EFT interactions with corresponding 2BC
    considered in this work (see Methods for details). The horizontal
    hatched and shaded areas are quenching factors obtained from the
    shell-model analyses of $\beta$-decays in the $p$-shell
    \cite{chou1993} and the $pf$-shell \cite{martinez1996},
    respectively.}  \label{Ikeda}
\end{figure}

The Ikeda sum-rule \citeapp{ikeda1963} (the difference between the
total integrated $\beta^-$ and $\beta^+$ strengths for Gamow-Teller
transitions mediated by $\bm {\sigma \tau^{\pm}} $) gives the
model-independent result $3(N - Z)$. Figure~\ref{Ikeda} shows our
coupled-cluster results for the quenching factor $Q \equiv q_{\rm
  2BC}^2 = (S^- - S^+)/ [3(N -Z)]$ for $^{14}$O, $^{48}$Ca, and
$^{90}$Zr obtained for the family of EFT Hamiltonians and
corresponding 2BC considered in this work (see Methods for
details). The total integrated Gamow-Teller strengths $S^{\pm} $ are
calculated as a ground-state expectation value by inserting a complete
set of intermediate $1^+$ states in the daughter nucleus using the
coupled-cluster method (see Methods for details).  We note that the
Ikeda sum-rule is trivially fulfilled if one employs the standard
Gamow-Teller operator $\bm {\sigma\tau}^\pm$.  Thus, the quenching
factor we obtain for the Ikeda sum-rule is entirely due to 2BC.

In summary, we obtain the quenching factors $q_{\rm 2BC}=0.90-0.94$
for $^{14}$O, $q_{\rm 2BC}= 0.80-0.86$ for $^{48}$Ca, and $q_{\rm
  2BC}=0.72-0.80$ for $^{90}$Zr, respectively. Our predictions for the
quenching factors are consistent with the shell-model analyses of
quenched $\beta$-decays in the $p$-shell \cite{chou1993}, $sd$-shell
\citeapp{wildenthal1983}, and the $pf$-shell \cite{martinez1996}. For
$^{48}$Ca our results for $S^-$ are in the range $18.5 -22.6$ and is
slightly larger than the measurement $S^- = 15.0 \pm 2.2$~\citeapp{yako2009}.
The measured strength in $^{90}$Zr is $S^- = 28.0 \pm 1.6$~\cite{wakasa1997}
and consistent with our calculated range of $21.5-29.6$.

\subsection{Role of 2BC and induced many-body terms in light nuclei}

\begin{figure}[h!]
  \includegraphics[width=1.0\columnwidth]{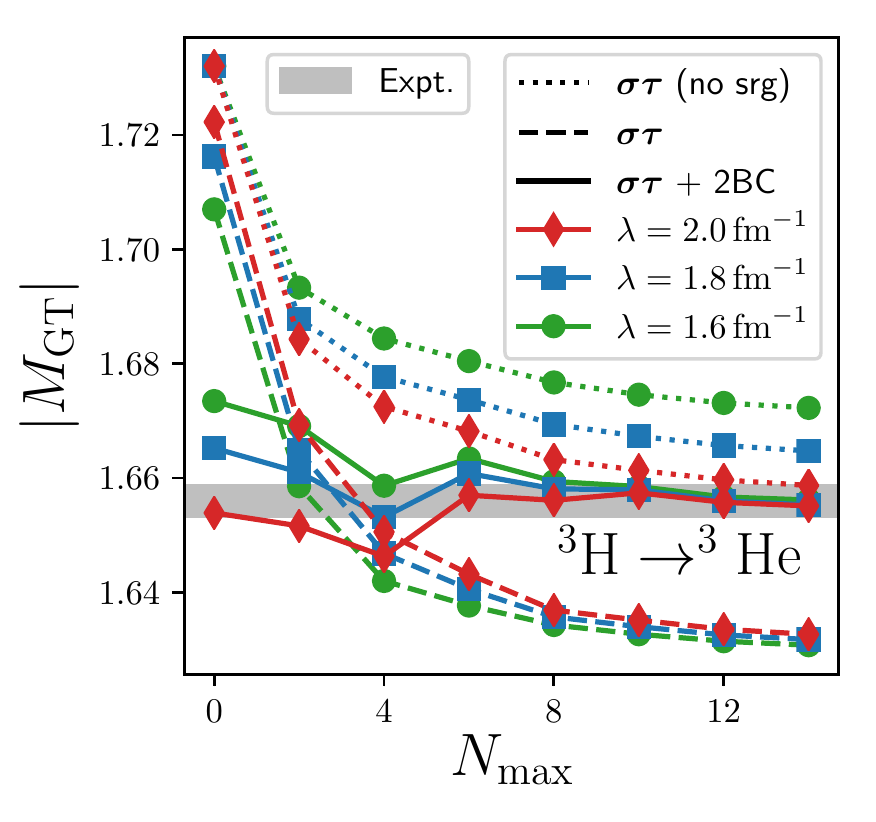}
  \includegraphics[width=1.0\columnwidth]{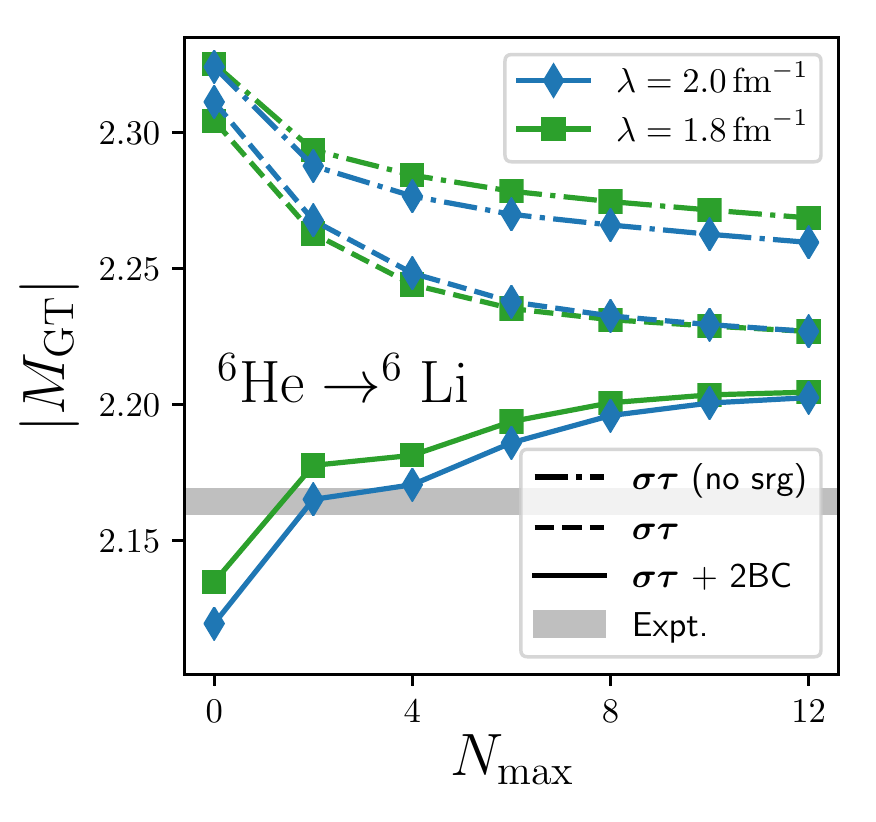}
  \caption{Convergence of the $^3$H$\rightarrow ^3$He (top panel) and
    $^6$He$\rightarrow ^6$Li (bottom panel) Gamow-Teller matrix
    elements with respect to the NCSM basis size for three values of
    the SRG evolution parameter $\lambda$.  Dashed and full lines show
    results obtained with one-body only and one- plus two-body
    operators, respectively.  The NN-N$^4$LO+3N$_{\rm lnl}$
    interaction was used with both the interaction and transition
    operators consistently SRG evolved.  The dotted lines show results
    ontained with the same SRG evolved interaction using only the
    one-body operator without any SRG evolution.  The shadow bands
    represent the experimental values with their uncertainties.}
  \label{ncsm1b}
\end{figure}

Let us study light nuclei in more detail, because these have been
described in several publications~\citeapp{gazit2009,ekstrom2014},
\cite{pastore2017}.  Figure~\ref{ncsm1b} shows the convergence of
$^3$H$\rightarrow ^3$He and $^6$He$\rightarrow ^6$Li Gamow-Teller
matrix elements with the NN-N$^4$LO+3N$_{\rm lnl}$ interaction and
highligts the size of the SRG induced two-body terms as well as the
role of the 2BC. Omitting the SRG-induced two-body contribution of the
$\sigma \tau$ operator results in a matrix element dependence on the
SRG evolution parameter $\lambda$. This depencence is then removed by
the induced two-body terms (i.e., the three- and higher-body induced
terms are negligible) from the consistent SRG evolution that at the
same time reduce the absolute values of the matrix elements.  The
(SRG-evolved) 2BC then enhance or quench the Gamow-Teller matrix
elements in case of $^3$H$\rightarrow ^3$He and $^6$He$\rightarrow
^6$Li$_{\tfrac{1}{2}}$, respectively.

\begin{figure}[h!]
  \includegraphics[width=1.0\columnwidth]{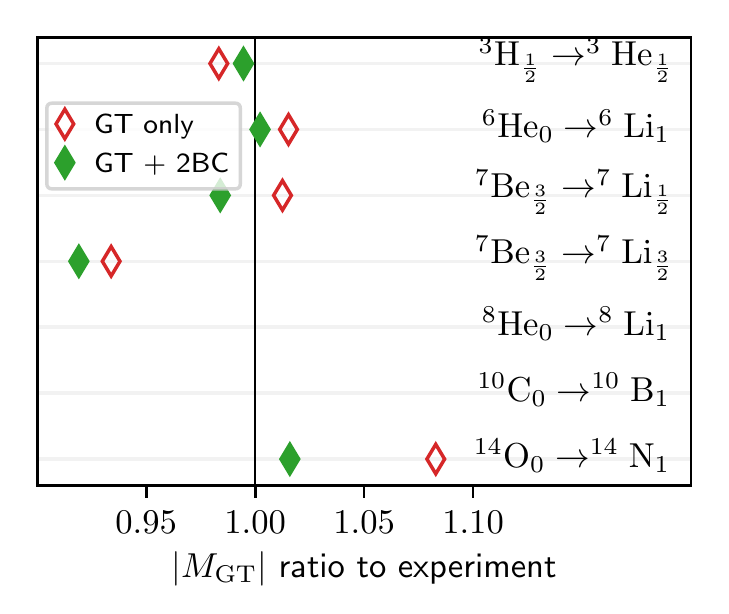}
  \includegraphics[width=1.0\columnwidth]{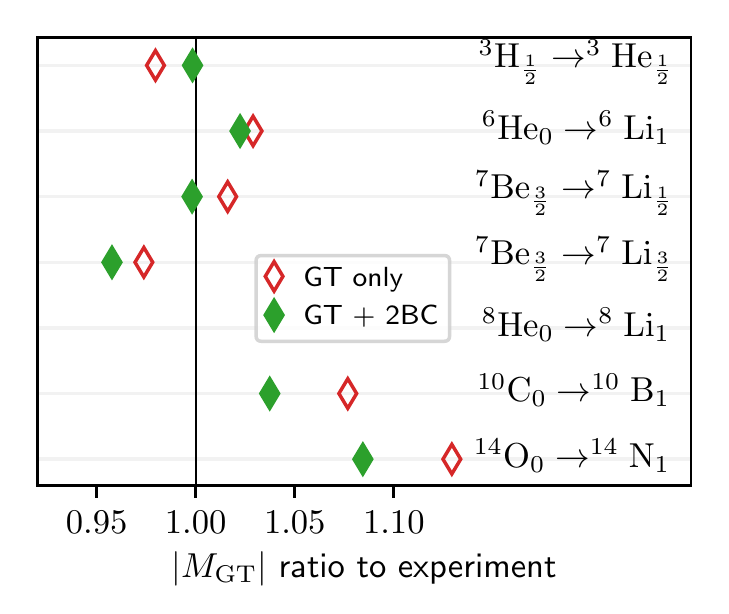}
  \caption{Theory-to-experiment ratio for the Gamow-Teller matrix
    elements of six strong transitions in light nuclei for the
    NNLO$_{\rm sat}$ (top) and NN-N$^3$LO+3N$_{\rm lnl}$ (bottom)
    interactions. The subscripts in the legend denote the total
    angular momenta of the parent and daughter states. All initial
    states are ground states. In the case of $^3$H$\rightarrow^3$He,
    $^6$He$\rightarrow^6$Li and $^7$Be$\rightarrow
    ^7$Li$_{\tfrac{3}{2}}$, the daughter nucleus is in its ground
    state, while the $^7$Be$\rightarrow ^7$Li$_{\frac{1}{2}}$ and
    $^{10}$C$\rightarrow ^{10}$B$_{1}$ are decays to the first excited
    state of the daughter nucleus, and the $^{14}$O$\rightarrow
    ^{14}$N$_{1}$ is a decay to the second excited state of
    $^{14}$N. Hollow symbols correspond to results obtained with the
    standard Gamow-Teller ${\bm \sigma}{\bm \tau}$ operator, and full
    symbols include 2BC. For NNLO$_{\rm sat}$ (top panel) the value
    for the $^{10}$C$\rightarrow ^{10}$B$_{1}$ Gamow-Teller transition
    are off-scale and not shown.}
  \label{ncsm2}
\end{figure}

Figure~\ref{ncsm2} shows theory-to-experiment ratios for strong
Gamow-Teller transitions in light nuclei for the NNLO$_{\rm sat}$ and
NN-N$^3$LO+3N$_{\rm lnl}$ interactions and corresponding 2BC. For
these interactions, the 3N forces and 2BC were not constrained to 
reproduce the triton half-life.

\subsection{Role of 2BC in $sd$- and $pf$-shell nuclei}
\begin{figure}[ht]
  \includegraphics[width=1.0\columnwidth]{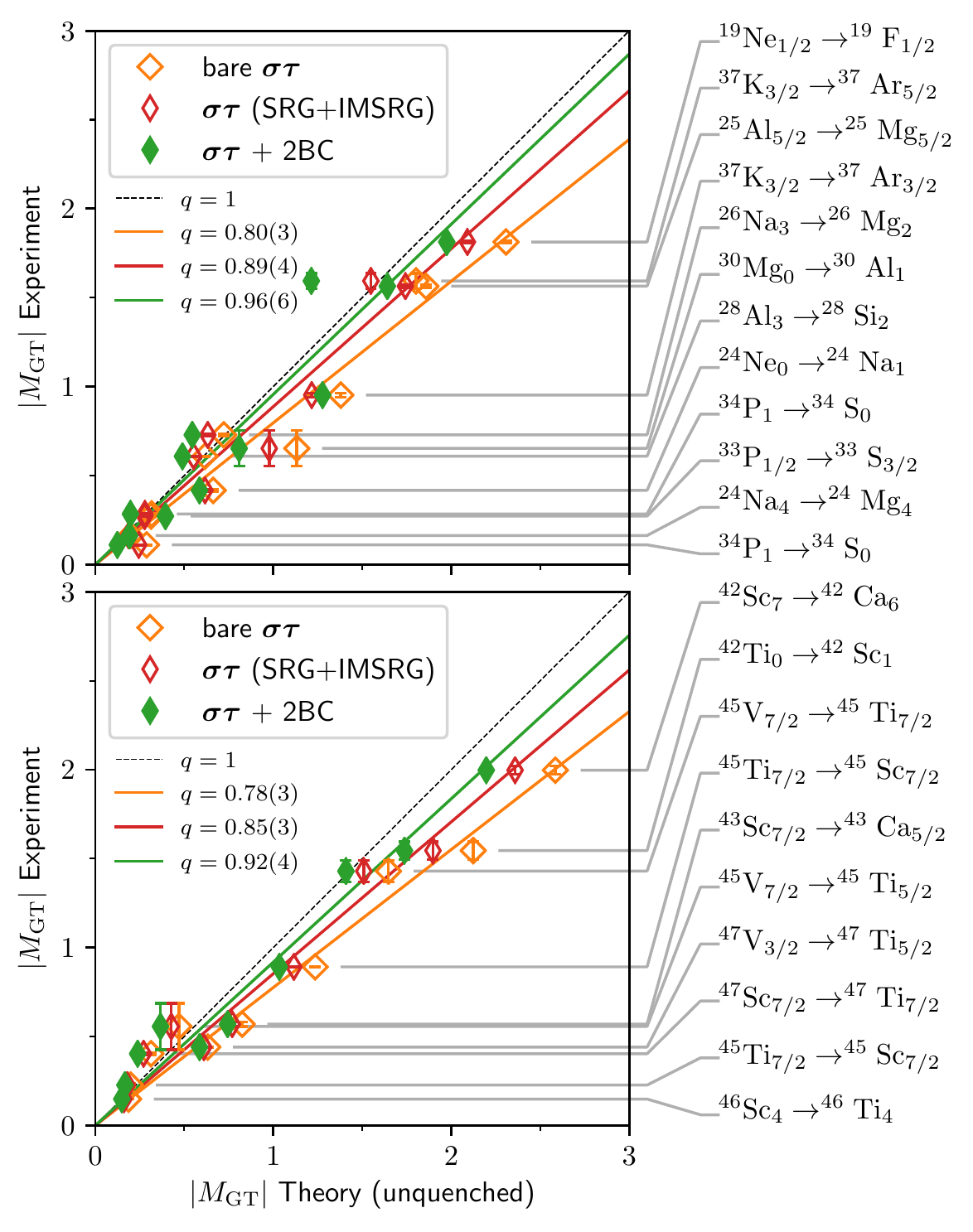}
  \caption{Comparison of experimental and theoretical Gamow-Teller
    matrix elements for medium-mass nuclei in the $sd$-shell (top
    panel) and lower $pf$-shell (bottom panel) for the
    NN-N$^4$LO+3N$_{\rm lnl}$ interaction. The theoretical results
    were obtained using (i) a bare Gamow-Teller operator ${\bm
      \sigma}{\bm \tau}$ (no SRG evolution), (ii) a ${\bm \sigma}{\bm
      \tau}$ operator consistently evolved with the Hamiltonian by SRG
    and IMSRG, and (iii) a consistently-evolved Gamow-Teller operator
    including 2BC. All expecation values are taken between the same
    VS-IMSRG wave functions. The linear fits show the resulting
    quenching factor $q$ given in the panels.}
  \label{sdshell2}
\end{figure}

Medium-mass nuclei have been key to indficate the quenching
puzzle~\cite{langanke1995,martinez1996,brown2006}. This makes it
important to present more details for these nuclei.
Figure~\ref{sdshell2} shows the empirical values of the Gamow-Teller
transition matrix elements versus the corresponding theoretical matrix
elements obtained with the standard Gamow-Teller ${\bm \sigma}{\bm
  \tau}$ operator (wide diamonds), with the consistently-evolved ${\bm
  \sigma}{\bm \tau}$ operator (narrow diamonds), and the full
calculation with the inclusion of 2BC (filled diamonds) for the
NN-N$^4$LO+3N$_{\rm lnl}$ interaction. Perfect agreement between
theory and experiment is denoted by the diagonal dashed line.  We see
that both the consistent evolution of the operator (which accounts for
correlations beyond the valence space) and the inclusion of 2BC
contribute significantly to the quenching effect, consistent with the
observations in Figs.~\ref{Sn100} and~\ref{fig:sn100_espm}.

\begin{figure}
\includegraphics[width=\columnwidth]{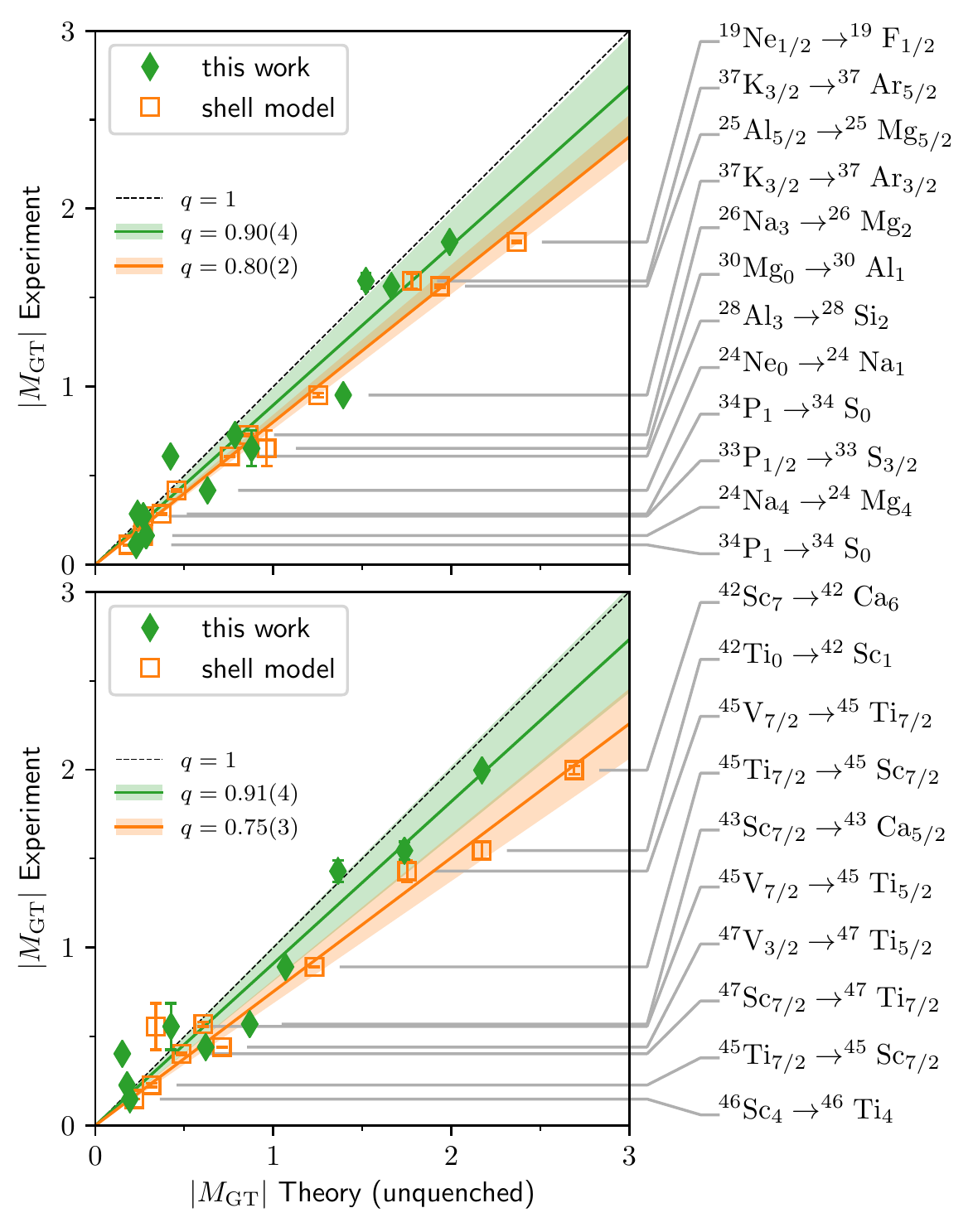}
\caption{Same as Fig.~\ref{sdshell}, except that we employ the
  NNLO$_{\rm{sat}}$ interaction in the VS-IMSRG calculations.}
\label{sdshellsat}
\end{figure}

In order to assess how dependent the quenching effect observed in
Fig.~\ref{sdshell} is on the choice of input
interaction, we repeat the calculation with the NNLO$_{\rm sat}$
interaction. The results, shown in Fig.~\ref{sdshellsat}, are
consistent with the findings in Fig.~\ref{sdshell}.

\subsection{Benchmarks between methods}

In this work we used three complentary methods -- the NCSM, the
coupled-cluster method, and the VS-IMSRG -- to compute nuclei in
different regions of the nuclear chart. For binding energies alone,
these methods have been benchmarked several times, see, e.g.,
Refs.~\cite{morris2017} and \citeapp{hergert2016}. Here, we also
benchmark these methods for Gamow-Teller decays. We focus on the
Gamow-Teller decays of the light nuclei $^8$He, $^{10}$C, and $^{14}$O
that exhibit considerable impact from 2BC. The coupled-cluster method
is limited to nuclei with closed subshells. In light nuclei, this
limits us to $^8$He and $^{14}$O, and in medium-mass nuclei we focus
on the Gamow-Teller decays of $^{34}$Si and $^{68}$Ni. Our
coupled-cluster calculations use the CCSDT-1 approximation. The
presented NCSM results are converged to within 3\% with respect to the
model-space size.

Table~\ref{tab:tab4b} shows the the transition in $^8$He to the first
$1_1^+$ state in $^8$Li using the NN-N$^4$LO +3N$_{\rm lnl}$
interaction. Here $M_{\rm GT}(\bm{\sigma\tau})$ denotes the matrix
element of the $\bm{\sigma\tau}$ operator, while $M_{\rm GT}$ includes
2BC. The NCSM and coupled-cluster triples results are within a few percent of
each other, while VS-IMSRG is within about 30\% of the NCSM.

\begin{table}[h]
  \caption{Gamow Teller (GT) transition strength in
    $^{8}$He to the first $1_1^+$ in $^{8}$Li for the NN-N$^4$LO +3N$_{\rm lnl}$ interaction
    calculated in the EOM-CCSDT-1, VS-IMSRG(2), and NCSM approaches.}
\begin{center}
\renewcommand{\arraystretch}{1.3}
\begin{tabular}{|l|c|c|}\hline
  Method   & \multicolumn{1}{|c|}{ $\vert M_{\rm GT}
  (\bm{\sigma\tau})\vert $} &\multicolumn{1}{c|}{$\vert M_{\rm
   GT}\vert$} \\\hline
 EOM-CCSD    &  0.45  & 0.48    \\
 EOM-CCSDT-1 &  0.42 &  0.45  \\
 VS-IMSRG(2) &  0.54 &  0.58 \\  
 NCSM        & 0.41 & 0.46  \\\hline
\end{tabular}
\renewcommand{\arraystretch}{1}
\end{center}
\label{tab:tab4b}
\end{table}

Table~\ref{tab:tab4} shows the transition in $^{14}$O to the second
$1_2^+$ state in $^{14}$Ni using the NNLO$_{\rm sat}$ and the
NN-N$^4$LO +3N$_{\rm lnl}$ interaction. The results obtained for the
different interactions and methods agree with each other to within
about 5\%, which is reasonable considering the level of approximation
(see methods section discussion of the VS-IMSRG above).

\begin{table}[h]
  \caption{Gamow Teller (GT) transition strength in
    $^{14}$O to the second $1_2^+$ in $^{14}$N for the NNLO$_{\rm
      sat}$ and NN-N$^4$LO +3N$_{\rm lnl}$ interactions
    calculated in the EOM-CCSDT-1, VS-IMSRG, and NCSM approaches.}
\begin{center}
\renewcommand{\arraystretch}{1.3}
\begin{tabular}{|l|c|c|c|c|}\hline
Interaction    &  \multicolumn{2}{c|}{NNLO$_{\rm sat}$} & \multicolumn{2}{c|}{ NN-N$^4$LO+3N$_{\rm lnl}$}   \\\hline
 Method      & $\vert M_{\rm GT} (\bm{\sigma\tau})\vert $ & $\vert
 M_{\rm  GT}\vert$& $\vert M_{\rm GT} (\bm{\sigma\tau})\vert $ &
 $\vert M_{\rm  GT}\vert$ \\\hline 
 EOM-CCSD       & 2.15    & 2.08               & 2.26   & 2.06   \\
 EOM-CCSDT-1    & 1.77    & 1.69               & 1.97   & 1.86   \\
 VS-IMSRG(2)    & 1.72    & 1.76               & 1.83   & 1.83   \\
 NCSM           & 1.80    & 1.69               & 1.86   & 1.78  \\\hline
\end{tabular}
\renewcommand{\arraystretch}{1}
\end{center}
\label{tab:tab4}
\end{table}

The benchmark for the Gamow-Teller decay of $^{10}$C between NCSM and
VS-IMSRG is shown in Tab.~\ref{tab:tab4c}. Both methods agree with
each other within 3\%.

\begin{table}[h]
  \caption{Gamow Teller (GT) transition strength in
    $^{10}$C to the first $1_1^+$ in $^{10}$B for the NN-N$^4$LO +3N$_{\rm lnl}$ interaction
    calculated in the VS-IMSRG(2) and NCSM approaches.}
\begin{center}
\renewcommand{\arraystretch}{1.3}
\begin{tabular}{|l|c|c|}\hline
  Method   & \multicolumn{1}{|c|}{ $\vert M_{\rm GT}
  (\bm{\sigma\tau})\vert $} &\multicolumn{1}{c|}{$\vert M_{\rm
   GT}\vert$} \\\hline
 VS-IMSRG(2) &  1.94 &  1.88 \\  
 NCSM        &  2.01 &  1.92  \\\hline
\end{tabular}
\renewcommand{\arraystretch}{1}
\end{center}
\label{tab:tab4c}
\end{table}

For medium-mass nuclei, the comparison between coupled-cluster methods
and VS-IMSRG is shown in Tab.~\ref{tab:tab4e}. The CCSDT-1 and
VS-IMSRG(2) results are within a few percent for the Gamow-Teller
decay of $^{64}$Ni, and within about 20\% and 25\% for the
Gamow-Teller decays of $^{34}$Si to the second and first excited $1^+$
states in $^{34}$P. We note that for the transitions in $^{34}$Si
there might be some mixing between the low-lying $1^+$ states in
$^{34}$P that is differently accounted for in the CCSDT-1 and
VS-IMSRG(2) approaches. To mitigate this effect we consider the
square root of the sum of the squared Gamow-Teller strengths to the first two $1^+$ states in
$^{34}$P (see e.g.~\cite{brown1985,martinez1996}), and obtain 
$[\sum M^2_{\rm GT}(\bm{\sigma\tau})]^{1/2}$ = 1.28 and 1.18,
and 
$[\sum M^2_{\rm GT}]^{1/2}$ = 1.0 and 0.89
for CCSDT-1 and VS-IMSRG(2), respectively.

\begin{table*}[h]
  \caption{Gamow Teller (GT) transition strengths in
    $^{34}$Si and $^{68}$Ni for the NN-N$^4$LO +3N$_{\rm lnl}$ interaction
    calculated in the EOM-CCSDT-1 and VS-IMSRG approaches.}
\begin{center}
\renewcommand{\arraystretch}{1.3}
\begin{tabular}{|l|c|c||c|c||c|c|}\hline
 & \multicolumn{2}{c||}{ $^{34}$Si$\rightarrow ^{34}$P$_{1_1^+}$} &
\multicolumn{2}{c||}{ $^{34}$Si$\rightarrow ^{34}$P$_{1_2^+}$} &
\multicolumn{2}{c|}{ $^{68}$Ni$\rightarrow ^{68}$Cu$_{1_1^+}$}  \\
 \multicolumn{1}{|c|}{Method}  & \multicolumn{1}{c}{$\vert M_{\rm
     GT}(\bm{\sigma\tau})\vert $} & 
\multicolumn{1}{c||}{$\vert M_{\rm
    GT}\vert$} &\multicolumn{1}{c}{$\vert M_{\rm GT}(\bm{\sigma\tau})\vert $} & \multicolumn{1}{c||}{$\vert M_{\rm
    GT}\vert$} &\multicolumn{1}{c}{$\vert M_{\rm GT}(\bm{\sigma\tau})\vert $} & \multicolumn{1}{c|}{$\vert M_{\rm
    GT}\vert$} \\\hline
 EOM-CCSD        & 0.61   & 0.28  & 1.32  & 1.14 &  0.60   & 0.46  \\
 EOM-CCSDT-1  & 0.56   & 0.28 & 1.15  & 0.96 &  0.42    & 0.35   \\
 VS-IMSRG(2)      & 0.78   & 0.52 & 1.00  & 0.85 & 0.42   & 0.36  \\\hline
\end{tabular}
\renewcommand{\arraystretch}{1}
\end{center}
\label{tab:tab4e}
\end{table*}

\subsection{Convergence of excited states and Gamow-Teller transitions}

\begin{figure}[ht]
  \includegraphics[width=1.0\columnwidth]{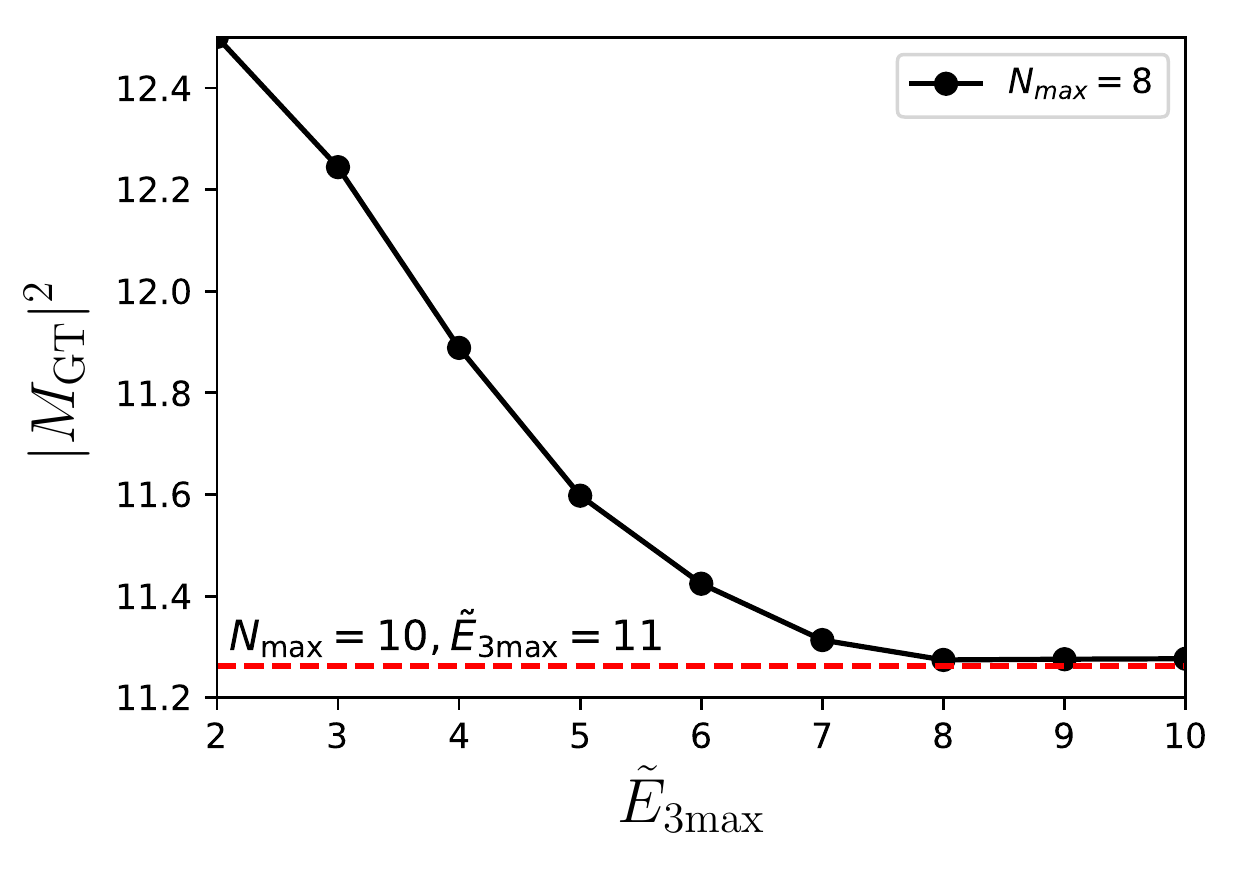} \caption{
  Convergence of the EOM-CCSDT-1 calculation of the
  Gamow-Teller transition of $^{100}$Sn with respect to the active
  space truncation $\tilde{E}_{\rm 3max}$ of the three-particle-three-hole
  excitations in the EOM-CCSDT-1 approximation for $N_{\rm max} = 8$
  for the 1.8/2.0 (EM) interaction.  The horizontal line shows the
  converged $N_{\rm max} =10, \tilde{E}_{\rm 3max}=11$
  result.}  \label{fig:Mgt_conv}
\end{figure}

Gamow-Teller decays often involve excited states of the daughter
nucleus.  For this reason we present deatils regarding the quality of
our calculations of excited states.  Figure ~\ref{fig:Mgt_conv} shows
the convergence of the Gamow-Teller transition of $^{100}$Sn with
respect to the active space truncation
$\tilde{E}_{pqr}=\tilde{e}_p+\tilde{e}_q+\tilde{e}_r<\tilde{E}_{\rm 3
  max}$ in the three-particle-three-hole excitations in EOM-CCSDT-1
for $N_{\rm max} = 8$ for the 1.8/2.0 (EM) interaction. By comparing
with the converged $N_{\rm max} =10, \tilde{E}_{\rm 3 max}=11 $ result
we see that the Gamow-Teller transitions is converged with respect to
both the model-space truncation $N_{\rm max}$ and the active space
truncation $\tilde{E}_{pqr}$. This new truncation allows for
accelerated convergence with minimal configurations both in ground-
and excited-state calculations.  Even for the hardest interaction,
NNLO$_{\rm sat}$, we find that the result is converged at the $1\%$
level for Gamow-Teller transitions for truncation $N_{\rm max} =10,
\tilde{E}_{\rm 3max}=11$.

\begin{figure}[ht]
  \includegraphics[width=1.0\columnwidth]{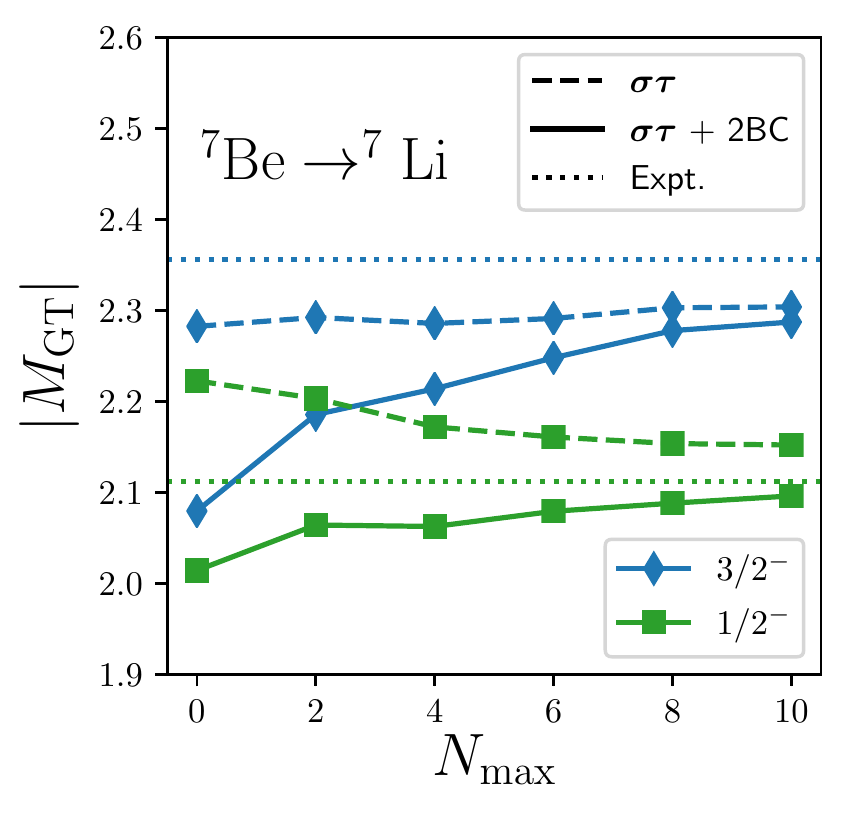} 
\caption{Convergence of the  $^7$Be$\rightarrow ^7$Li$_{\tfrac{3}{2}}$ 
and  $^7$Be$\rightarrow ^7$Li$_{\tfrac{1}{2}}$  Gamow-Teller matrix elements 
with respect to the NCSM basis size. Dashed and full lines show results 
obtained with one-body only and one- plus two-body operators, respectively. The dotted lines represent 
the experimental values. The NN-N$^4$LO+3N$_{\rm lnl}$ interaction was used with both the interaction 
and transition operators consistently SRG evolved.} \label{fig:Energy_conv2}
\end{figure}

\begin{figure}[ht]
  \includegraphics[width=1.0\columnwidth]{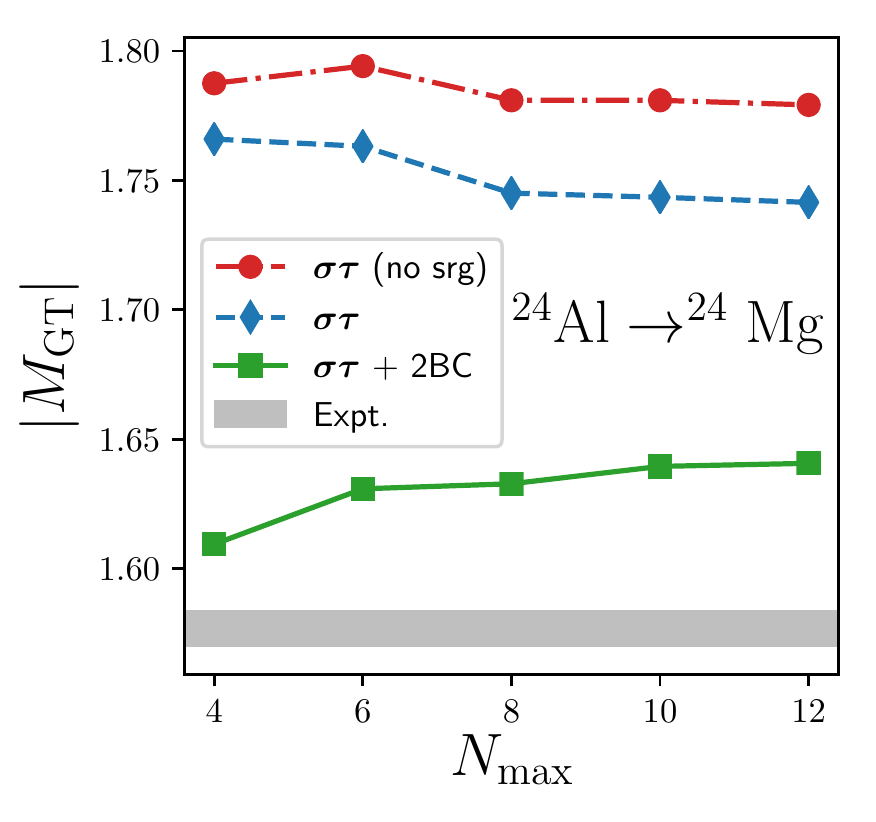}
\caption{Convergence of the Gamow-Teller matrix element for $^{24}$Al$\rightarrow^{24}$Mg
         calculated with the VS-IMSRG, as a function of the number of harmonic oscillator
          shells used.}  \label{fig:Energy_conv4}
\end{figure}

Figure~\ref{fig:Energy_conv2} shows the convergence of the
$^7$Be$\rightarrow ^7$Li$_{\tfrac{3}{2}}$ and $^7$Be$\rightarrow
^7$Li$_{\tfrac{1}{2}}$ Gamow-Teller matrix elements with respect to
the NCSM basis size. By looking at the convergence trend with
increasing model-space size it is conceivable that the effect of 2BC
could slightly enhance (instead of quench) the Gamow-Teller matrix
element for the $^7$Be$\rightarrow ^7$Li$_{\tfrac{3}{2}}$ transition,
bringing this result closer to those of
Ref.~\cite{pastore2017}. Finally Fig.~\ref{fig:Energy_conv4} shows the
convergence of the Gamow-Teller matrix element for the transition
$^{24}$Al$\rightarrow^{24}$Mg with increasing model-space size
calculated with the VS-IMSRG approach.

\begin{figure}[ht]
  \includegraphics[width=1.0\columnwidth]{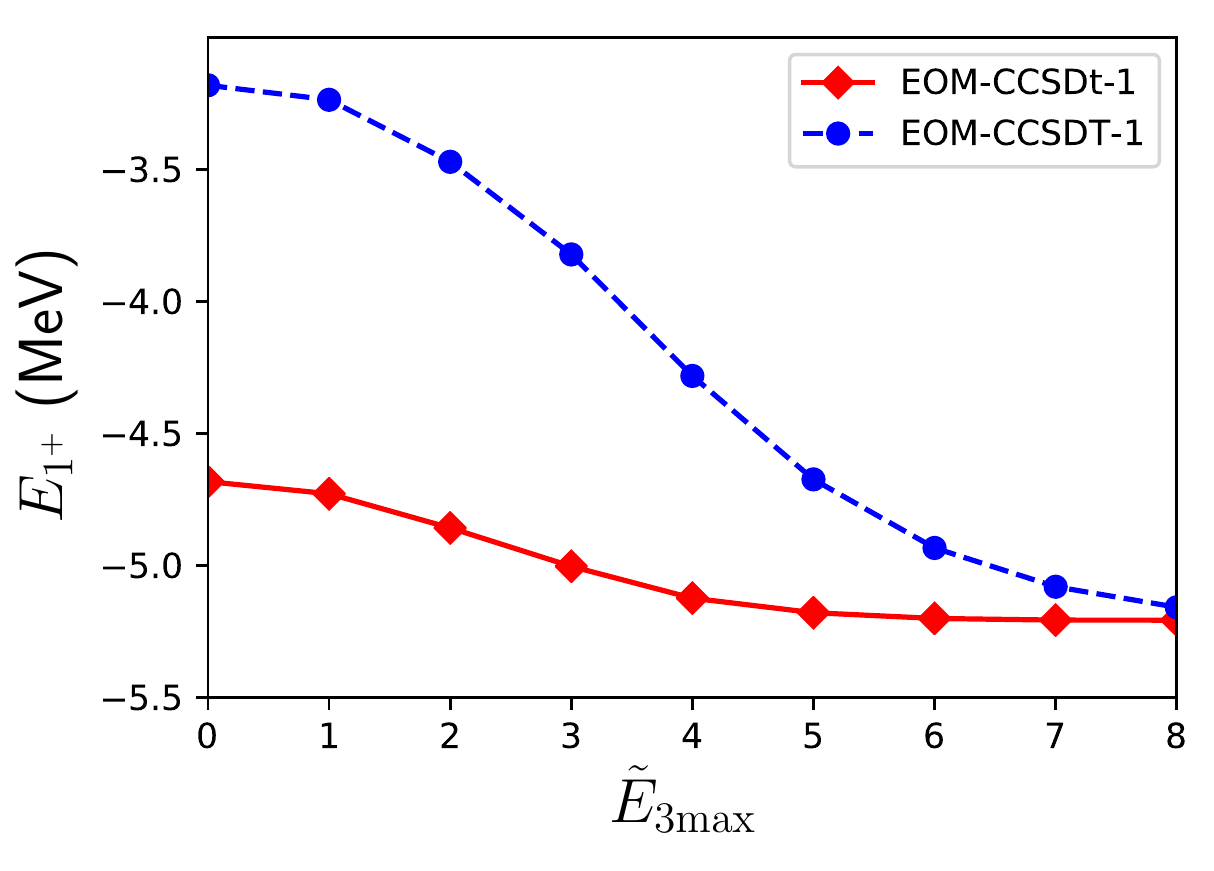} \caption{
    Convergence of the first excited $1^+$ state in $^{100}$In with
    respect to the ground-state energy of $^{100}$Sn for increasing
    $\tilde{E}_{\rm 3max}$ using the EOM-CCSDT-1 (blue dashed line)
    and EOM-CCSDt-1 (red solid line) approaches. Here the
    single-particle model-space was given by $N_{\rm max} = 8$, and we
    employed the 1.8/2.0 (EM) interaction.}  \label{fig:Energy_conv}
\end{figure}

Figure~\ref{fig:Energy_conv} shows the convergence of excited states
in the daughter nucleus $^{100}$In using the EOM-CCSDt-1 and the
EOM-CCSDT-1 approaches with $\tilde{E}_{\rm 3max}$. We clearly see a
dramatic acceleration in convergence with the EOM-CCSDt-1 approach
compared to the EOM-CCSDT-1 approach with respect to
$\tilde{E}_{\rm 3max}$.

\subsection{Fermi gas model}

The Fermi-gas model is one of the simplest approaches to atomic
nuclei. To throw further light onto the quenching factors we obtained
from 2BC, we compare the results of our sophisticated many-body
calculations with estimates from this simple model.  In the Fermi-gas
model of Ref.~\citeapp{menendez2011} the quenching due to 2BC is
\begin{equation}
q_{\rm 2BC}\approx 1 - {\rho \over F_\pi^2}\left[-{c_D\over 4 g_A \Lambda_\chi} +
{I(\rho)\over 3} \left(2c_4-c_3+{1\over 2m}\right)\right] \,,
\label{gren}
\end{equation}
with 
\begin{equation}
I(\rho) =1-{3m_\pi^2 \over k_F^2} + {3m_\pi^3 \over k_F^3} \arctan(k_F/m_\pi)
\label{gren2}
\end{equation}
and 
\begin{equation}
\rho = {2k_F^3 \over 3\pi^2} .
\label{gren3}
\end{equation}
Here, we employed the relationship between 2BC and 3N forces with the
correct factor $-1/4$ for the short range term proportional to $c_D$
as given in Ref.~\citeapp{krebs2017}.  In Eqs.~(\ref{gren}) to
(\ref{gren3}), $\rho$ is the density, $k_F$ the Fermi momentum, and
$I(\rho)$ is a function of $\rho$. We have $\rho = 0.16$~fm$^{-3}$,
$k_F = 1.33$~fm$^{-1}$, and $I(\rho) \approx 0.65 $ at nuclear matter
saturation, $F_\pi = 92.4$~MeV the pion-decay constant, $m = 939$~MeV
the nucleon mass, $g_A = 1.27$, and $\Lambda_\chi = 700$~MeV. The
pion-nucleon couplings $c_{3}$ and $c_4$ yield a value $2c_4-c_3$ of
about 11 to 15~GeV$^{-1}$ for a wide range of EFT interactions (see
Table~\ref{tab:tab1} and Methods for details).  These contributions
have been considered early on (see, e.g., Ref.~\cite{towner1987}), and
in our EFT-based calculations give the dominant contributions from 2BC
to the quenching. The 2BC contribution proportional to $c_D$ is set by
shorter-range 3N forces and is found to be small, which counters the
quenching from the long-range part of the 2BC. We note that acceptable
nuclear saturation favors values for $c_D$ that are positive and of
order one, and we have $0.7\lesssim c_D \lesssim 1.3$ for the EFT
interactions we employ (except for the NN-N$^4$LO+3N$_{\rm lnl}$ and
2.0/2.0 (PWA), see Table~\ref{tab:tab1} below and Methods).

We can attempt to relate the results displayed in Fig.~\ref{Sn100} and
Table~\ref{tab:tab2} to the Fermi gas model. For this purpose we
show in Table~\ref{tab:tab1} the relevant low-energy couplings that
enter Eq.~(\ref{gren}), both for the EFT interactions used in this
work and for other interactions that yielded little contributions from
2BC.  We note that the contributions from the pion-exchange 2BC
(proportional to $2c_4-c_3$) vary less than those from the
short-ranged 2BC (proportional to $c_D$).

\begin{table}[ht]
\caption{Values of the low-energy coupling $c_D$, the combination
$2c_4-c_3$ of pion-nucleon couplings, and the value of $\Lambda_\chi$
that enters the simple estimate~(\ref{gren}) of the
renormalization of the axial-vector coupling due to 2BC. The first
seven EFT interactions are used in this work, and the latter are other
interactions (for details see the references given).}
\begin{center}
\renewcommand{\arraystretch}{1.3}
\begin{tabular}{|l|D{.}{.}{4}|D{.}{.}{2}|c|c|}\hline
  Interaction & c_D & 2c_4-c_3 & $\Lambda_\chi$ [GeV] & Ref. \\ \hline 
  NNLO$_{\rm sat}$ & 0.817 &11.46& 0.7 & \cite{ekstrom2015} \\
  NN-N$^4$LO +3N$_{\rm lnl}$& -1.8 &13.88 & 0.7 & \\
  NN-N$^3$LO +3N$_{\rm lnl}$& 0.7
  &14.0&0.7&\cite{leistenschneider2017}\\ 1.8/2.0 (EM) & 1.264
  &14.0&0.7&\cite{hebeler2011}\\ 2.0/2.0 (EM) & 1.271
  &14.0&0.7&\cite{hebeler2011}\\ 2.2/2.0 (EM) & 1.214
  &14.0&0.7&\cite{hebeler2011}\\ 2.0/2.0 (PWA) &-3.007
  &12.7&0.7&\cite{hebeler2011}\\\hline  Pastore 500 &-1.847
  &14.0&1.0&\cite{pastore2017}\\ Pastore 600 &-2.03
  &14.13&1.0&\cite{pastore2017}\\ 
  Ekstr\"om 450 & 0.0004
  &13.22&0.7&\citeapp{ekstrom2014}\\ Ekstr\"om 500 & 0.0431
  &12.50&0.7&\citeapp{ekstrom2014}\\ Ekstr\"om 550 & 0.1488
  &11.71&0.7&\citeapp{ekstrom2014}\\ \hline
\end{tabular}
\renewcommand{\arraystretch}{1}
\end{center}
\label{tab:tab1}
\end{table}

\begin{figure}[ht] 
  \includegraphics[width=1.0\columnwidth]{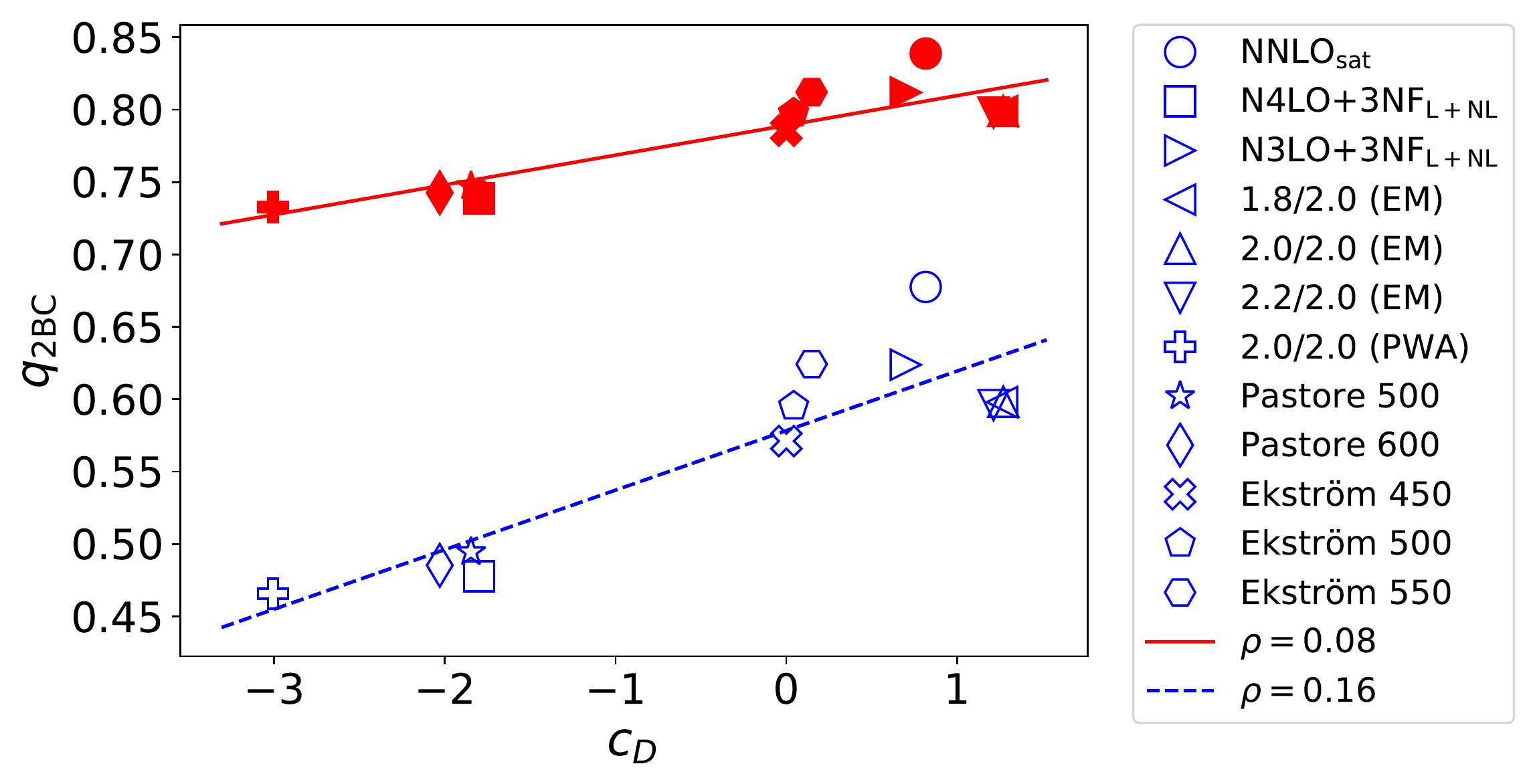}
  \caption{Quenching factor $q_{\rm 2BC}$ obtained from Eq.~(\ref{gren}) for the
    EFT interactions considered in this work as well as from
    Refs.~\protect\citeapp{ekstrom2014}, \protect\cite{pastore2017} (see
    Table~\ref{tab:tab1}), at two different densities $\rho =
    0.08$~fm$^{-3}$ and $\rho=0.16$~fm$^{-3}$.}
 \label{fig:fig-quench}
\end{figure}

Figure~\ref{fig:fig-quench} shows the quenching factor based on
Eq.~(\ref{gren}) for interactions as indicated (compare to
Table~\ref{tab:tab1} for references). The full red symbols are for a
density $\rho=0.08$~fm$^{-3}$ (typical for the nuclear surface), and
the full red line shows a linear fit to the points. The hollow blue
symbols and dashed line show the corresponding results at the
saturation density $\rho=0.16$~fm$^{-3}$. The contributions of 2BC are
proportional to the density $\rho$.

\begin{figure}[htb]
  \includegraphics[width=1.0\columnwidth]{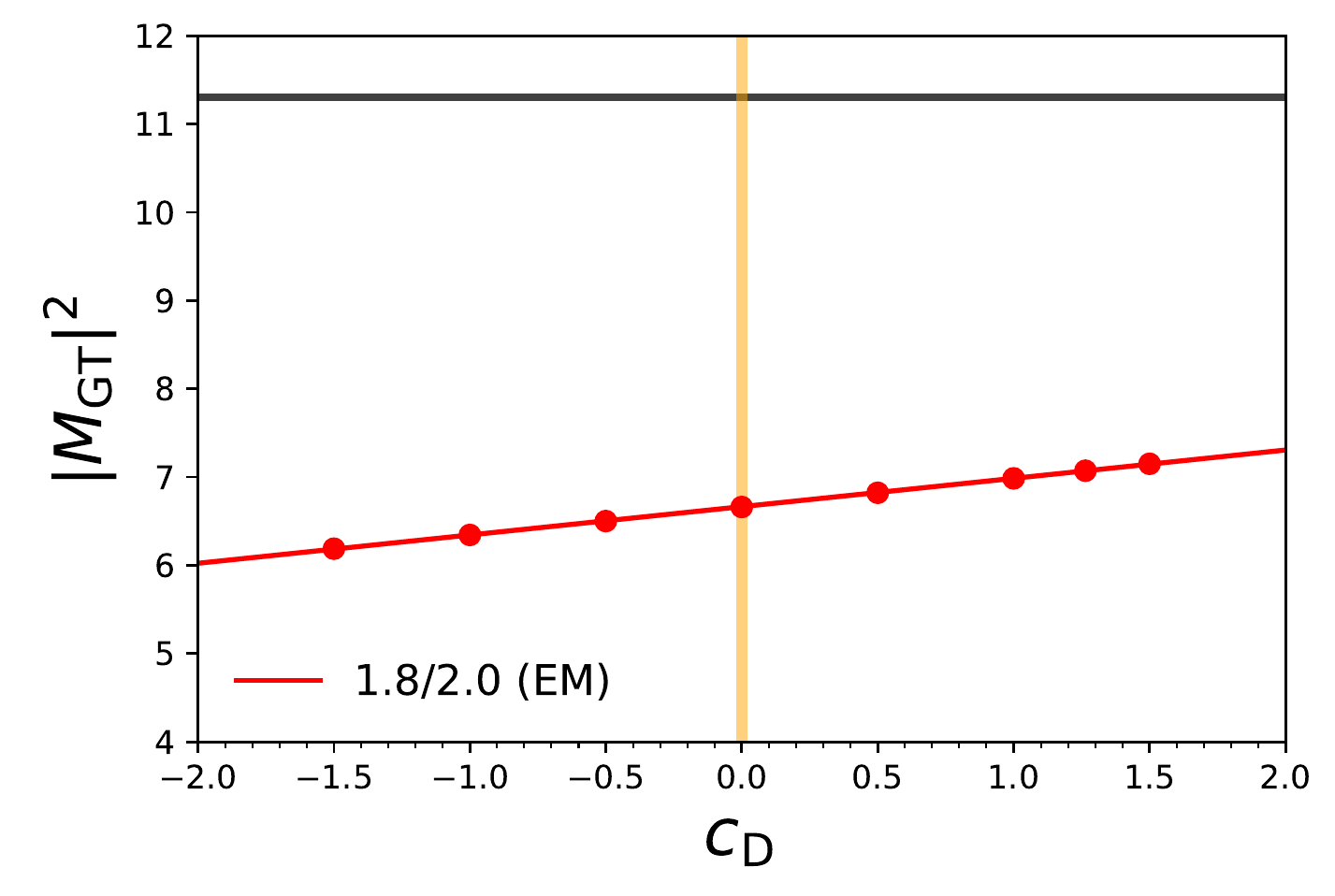} \caption{
    Gamow-Teller strength $|M_{\rm GT}|^2$ in $^{100}$Sn as a function
    of the low-energy coupling $c_D$ for the 1.8/2.0 (EM)
    interaction. The horizontal black line gives the result for the
    Gamow-Teller strength for the standard Gamow-Teller operator ($\bm
    \sigma \bm \tau$) only. As can been seen, the quenching from the
    pion-exchange 2BC alone ($c_D=0$) is $q_{\rm 2BC} = 0.77$, and for the full
    2BC ($c_D = 1.264$) it is $q_{\rm 2BC} = 0.79$. Thus the majority of the
    quenching from 2BC comes from the long-range pion-exchange part.}
  \label{fig:fig-quench2}
\end{figure}

In Fig.~\ref{fig:fig-quench2} we show the Gamow-Teller transition
strength in $^{100}$Sn as a function of the low-energy coupling $c_D$
for the interaction 1.8/2.0 (EM) computed in the charge-exchange
EOM-CCSDT-1 approach. For $c_D=0$ the extracted quenching factor is $q_{\rm 2BC}
= 0.77$, and for the full 2BC ($c_D = 1.264$) it is $q_{\rm 2BC} = 0.79$. This
demonstrates that the majority of the quenching from 2BC comes from
the long-range pion-exchange 2BC, while the short-range part gives a
smaller contribution.

\begin{figure}[htb] 
  \includegraphics[width=1.0\columnwidth]{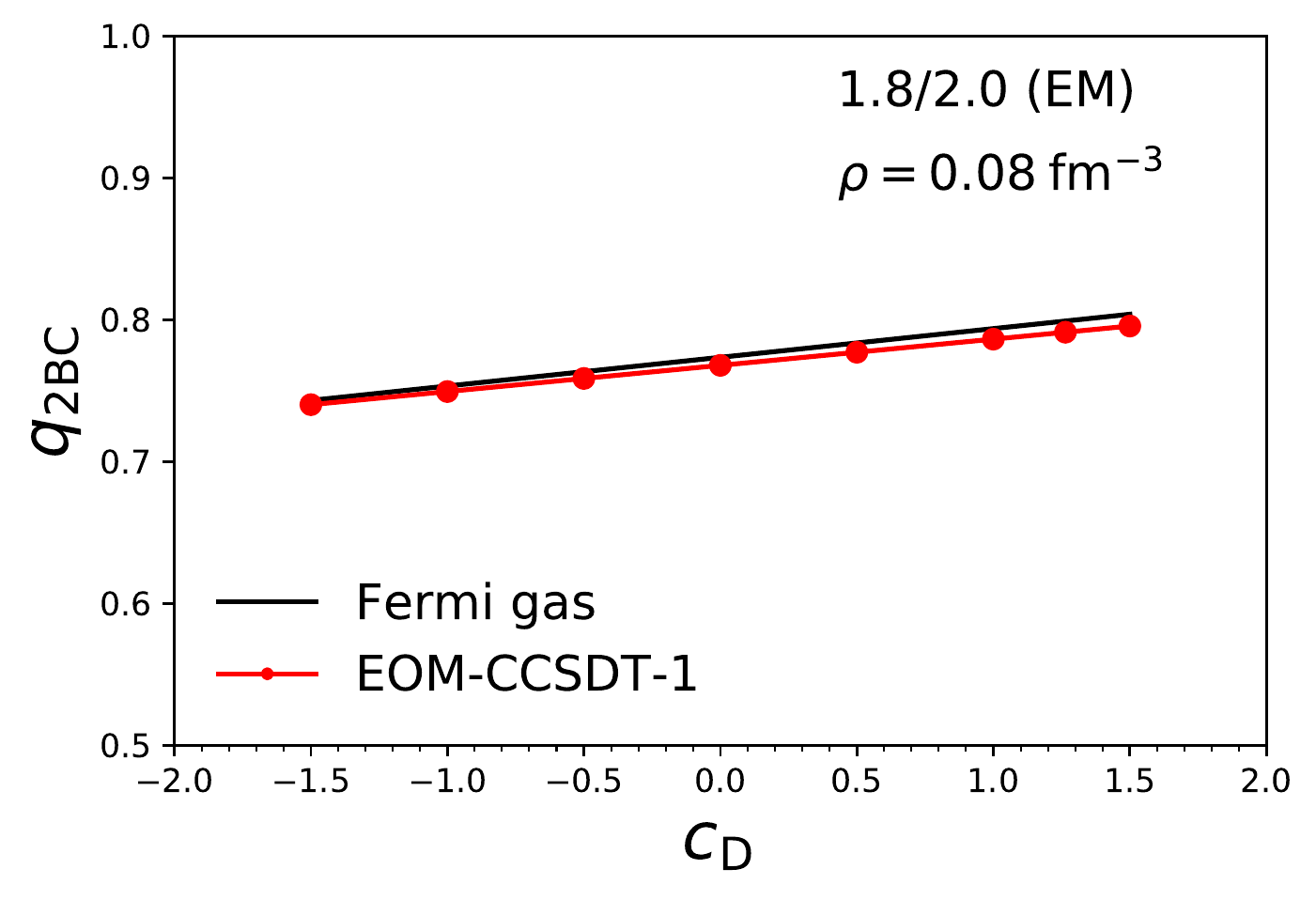} \caption{
    Quenching factor $q_{\rm 2BC}$ as a function of the low-energy coupling
    $c_D$ for the 1.8/2.0 (EM) interaction. The black line gives the
    quenching factor obtained with the Fermi-gas model and the red
    line gives the quenching factor obtained in the EOM-CCSDT-1
    method.}
 \label{fig:fig-quench3}     
\end{figure}

This accounting is further analyzed in
Fig.~\ref{fig:fig-quench3}. Here, the quenching factor for the
Gamow-Teller transition in $^{100}$Sn is shown as a function of $c_D$
for the Fermi-gas model at a density $\rho = 0.08~\rm{fm}^{-3}$ and
EOM-CCSDT-1 using the same interaction. We observe that the simple
interpretation of quenching in terms of the Fermi-gas model is
validated in this nucleus because both approaches yield the same linear
trend and similar values for the quenching as a function of the
low-energy coupling $c_D$.

\end{document}